\documentclass{book}
\usepackage[contrib,lang=american]{ems-book} 

\usepackage{physics} 
\begin{document}
\mainmatter

\title{A review of mechanistic and data-driven models of terrorism and radicalization}
\titlemark{Models of terrorism and radicalization}

\emsauthor{1}{Yao-Li Chuang}{Y. L. ~Chuang}
\emsauthor{2}{Maria R. D'Orsogna}{M. R.~D'Orsogna}


\emsaffil{1}{Postal address \email{ylch07@gmail.com}}
\emsaffil{2}{Postal address \email{dorsogna@csun.edul}}

\classification
{Game theory, economics, finance, and other social and behavioral sciences}

\keywords{mathematical modeling, data-analysis, radicalization, terrorism}


\begin{abstract}
The rapid spread of radical ideologies in recent years has led to a worldwide string of terrorist attacks. 
Understanding how extremist tendencies germinate, develop, and drive individuals to action 
is important from a cultural standpoint, but also to help formulate response and prevention strategies.  
Demographic studies, interviews with radicalized subjects, analysis of terrorist databases,
reveal that the path to radicalization occurs along progressive steps, where
age, social context and peer-to-peer exchange of extremist ideas play major roles. 
Furthermore, the advent of social media has offered new channels of communication, 
facilitated recruitment, and hastened the leap from mild discontent to unbridled fanaticism.
While a complete sociological understanding of the processes and circumstances that lead to full-fledged extremism is still lacking, 
quantitative approaches, using modeling and data analyses, can offer useful insight. 
We review some approaches from statistical mechanics,  applied mathematics, data science, that can help 
describe and understand radicalization and terrorist activity.  Specifically, 
we focus on compartment models of populations harboring extremist views, continuous time models for age-structured radical
populations, radicalization as social contagion processes on lattices and social networks,  adversarial evolutionary games coupling
terrorists and counter-terrorism agents, and point processes to study the spatio-temporal clustering of terrorist events. 
We also present recent applications of machine learning methods on open-source terrorism databases. 
Finally, we discuss the role of institutional intervention and the stages at which de-radicalization strategies might be most effective. 
\end{abstract}

\makecontribtitle


\section{Introduction}
\label{sec:1}
The terrorist attacks of September $11^{th}$ 2001 destroyed lives and buildings, inflicting pain on many communities. 
More than twenty years later, they have also profoundly changed our lives, from the mundane aspects of boarding an airplane, to the definition of
war, peace, tolerance. In their aftermath, it is no longer shocking to view a line of code, a truck, a pair of shoes, an aircraft as part of a sinister threat.
At the same time, in the name of security and safety, we have witnessed the erosion of some of our civil liberties, and we have come to accept that 
the privacy of our email exchanges, computer log-ins, financial transactions, and travel records may no longer be sacrosanct. One 
of the most troubling aspects of 
the September $11^{th}$ events is that they instilled fear in the population, contributing to greater political polarization and divisiveness
that continue to this day. 
Of course, this is the goal of terrorism, which in a nutshell is a political and psychological strategy to weaken the ``enemy'' from 
within \cite{HUD02, SPI09}.

\begin{figure}[t]

\includegraphics[width=4.7in]{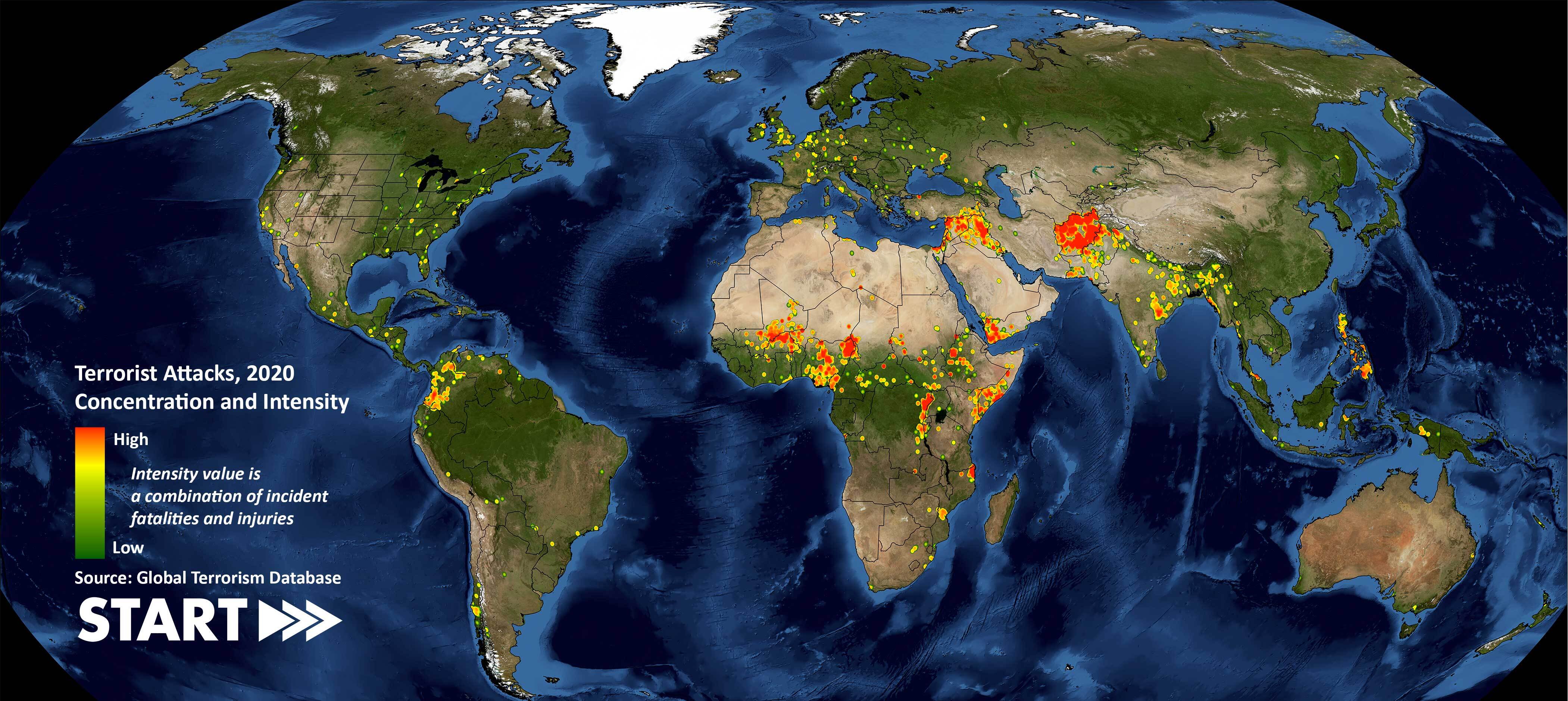}
\caption{Terrorist attacks between 1970-2020, from the Global Terrorist Database 
(GTD) as compiled by the National Consortium for Study of Terrorism and Responses to Terrorism
(START). The GTD details more than 200,000 events including
bombings, assassinations, and kidnappings. Data is entered upon reviewing
news articles, court reports and other sources. Among the most notable groups
are: the FARC in Colombia, Shining Path in Peru,
the FMLN in El Salvador; 
the Red Brigades in Italy, 
the Red Army Faction in Germany, 
the PIRA in Northern Ireland and 
the ETA in the Basque country; 
Boko Haram in Nigeria and Al-Shabaab in Somalia; 
Hamas, Hezbollah,  ISIS, Al -Qaeda and the Taliban in the Middle East; 
the LTTE and the NPA in South and Southeast Asia,
 as well as several groups in Jammu and Kashmir.}
\label{start}
\end{figure}

While on an unprecedented scale, the events of September $11^{th}$ are not isolated: terrorism, radicalization and the glorification of violence 
are worldwide phenomena that aim to disrupt civilian life and threaten security \cite{DEK06}. 
Fig.\,\ref{start} shows global terrorist attacks 
between 1970 and 2015 as compiled by the National Consortium for Study of Terrorism and Responses to Terrorism \cite{STA18}.
As can be seen, many corners of the world are hot-spots for terrorist activity, inspired and driven by a wide spectrum of motivations or ideologies
that can be religious, political, nationalist or subversive. Extremists exploit traditional and, in an increasingly globalized world, 
online tools to disseminate propaganda, recruit new adherents, plot attacks \cite{DEA16}. They do so often unhindered by geographical boundaries.  The question of
how to guarantee an open society, freedom of expression, belief, movement while at the same time preventing violence, 
is an unresolved one, dating all the way back to classical antiquity \cite{YON56}.

Despite the many limitations imposed by the clandestine nature of extremist operatives,
recent years have seen great advances in the collection, data-mining and statistical analysis 
of terrorist and radical data, such as profile lists of known extremists, databases listing terrorist attacks worldwide, the dissecting 
of twitter accounts, text analysis, the cataloguing of relevant hashtags and 
applying social network analysis \cite{GUO07, GAR09, SUB13a, MEM09, MUL13, MAH10, DES13, ZEC16, DIN17, ELZ18}.
Delving into the data allows us to find features and patterns, to uncover 
intriguing irregularities, and spurs our curiosity. However, seldom does it lead to the identification
of the underlying mechanisms causing the observed trends.

Mathematical models offer an alternate path. Since they are based on observations that can be 
described by modifiable dynamical processes, initial conditions and parametric variables,  
mathematical models allow us to analyze the effects of individual inputs, the interplay between them,
the effects of possible feedbacks and behaviors at different time scales. 
By using these inputs as building blocks, one can gradually build a nuanced understanding of the complex system
at hand, identify emergent behavioral or spatio-temporal patterns, test intervention methods, identify the effects of constraints.
Of course, measurability and quantifiability remain an issue and one must never forget that although useful in offering
insight, no model can capture the full complexity of the real world.
A relatively new approach is to conduct laboratory experiments, whereby 
participants are placed in virtual scenarios and must interactively select actions to take. 
They are then polled on their intents and motivations \cite{REE13, ARC11}. 
Perhaps, the future lies in combining data-mining and other statistical analysis of compiled
records with model-building, using tools from physical, mathematical and behavioral sciences.

In this work, we review recent models of radicalization and terrorism where ODEs, statistical physics, point processes, 
game and network theory are used to understand the multiple aspects involved, and to identify possible prevention or amelioration strategies.
A considerable body of work on modeling societal dynamics
is already present in the literature \cite{CAS09, NOW09, WAT07}, with applications to opinion, rumor and voter dynamics 
\cite{ZAN02, GAL05, SOO05, GAL07, MAR12, JAV14, JAV15, JEF16},  
 crowd behavior \cite{BUN69, PAN94, FOT12},  
crime and conflict \cite{COH81, LIC92, EPS02, JON11, DOR13, MOR13, DOR15, MAR15},  
gang affiliation \cite{HEG11, SMI12, BAR13},  segregation \cite{SCH71, GUR74, YIZ04, CHU19}, 
drug use \cite{BOB07, MOL20, CHO22, GUP22, MAO24}, 
social unrest \cite{BUR78, AND97, CAD15, BER15, CAR22, PET25},  
recidivism and rehabilitation \cite{GRE79, MAL77, BER14, FAR16}, culture dynamics \cite{CAV81, CAV83, GRA11, GON12},  
human cooperation \cite{PER17, SHO10, SHO13, SAN05}, warfare dynamics \cite{LAN56, DIX10, STR11}. 
Here, we specifically focus on the radicalization of  interacting cohorts, the evolution of 
terrorist activity, and possible intervention strategies.

After providing a general background on terrorism and radicalization in Sec.\,\ref{background}, 
we present compartment models that describe the radicalization process within societies in Sec.\,\ref{compartment}.
We discuss age structured models in Sec.\,\ref{agestructure} and network models in Sec.\,\ref{networks}
that allow for more nuance in representing interactions among populations and terrorist groups. In Sec.\,\ref{gametheory}
we review game theoretic models where terrorist groups and counter-terrorist agencies pursue different
strategies to maximize their objectives. Section\,\ref{sec:self-exciting} is focused on sequences of terrorist events 
represented as self-exiting processes to uncover any spatio-temporal clustering among them. 
In Sec.\,\ref{data} we review the open-source databases that detail terrorist events, 
and discuss recent applications to machine-learning methods.  
Finally, we close in Sec.\,\ref{conclusions} with a brief discussion and a word of caution 
on the applicability of these results to the real-world. 

\section{Background}
\label{background}

A terrorist event does not occur out of the blue. Individuals who commit these acts have typically spent months or years ``radicalizing:''
reading fanatical sources, discussing with, or emulating, others with more extreme views, becoming proactive
and later orchestrating attacks with like-minded peers. Numerous sociological, psychological and economic studies
have attempted to understand the overall process of radicalization,  
through demographic studies, by interviewing subjects, and by analyzing the socio-economic backgrounds and \textit{modus operandi} of extremists
\cite{SAG04, BAK06, LOZ11, DEC12, AHM13, GIL15, SCH15, SCA16}. One example is the 
Profiles of Individual Radicalization in the United States (PIRUS),
a database of domestic Islamist, Far Left and Far Right radicalized
individuals \cite{PIR18}. Distinctions have been made to differentiate when ideological radicalization stays part of a belief system, however contorted, 
and when it leads to bloodshed \cite{BAR12, ASA16, JEN18}. 
What has emerged is that while some general trends can be outlined, there is no one single, 
key element that can explain why a person, or a group of people, 
choose to reject the \textit{status quo} and embrace violence \cite{POS06, GAL23, CAM24}. 

The making of a radical typically unfolds in a multi-step fashion and is strongly influenced by 
one's surroundings, past experiences and future prospects \cite{SIL03, VIC06, WRI06, SIL07, MCC08, SIL08}. 
De\-pending on the particular socio-geographical context, four or five main
steps have been identified: pre-radicalization (at risk), self-identification (susceptible),
indoctrination (moderate radical), commitment and ``jihadization" (full radicalization) \cite{SIL07, CRE11, NEO16}. 
More nuanced classifications, expanding these classes to eight states, have also been proposed \cite{COE18}; 
IVEE theory instead focuses on three main steps towards radicalization: individual vulnerability (IV), exposure (E), emergence (E)
 \cite{BOU10, BOU11, PEP20}.
The time-scales for individuals to advance through these hierarchies vary, but 
case studies conducted on sixty-eight American homegrown Al-Qaeda inspired radicals 
reveal that pre-radicalization takes four to five years, progressing through the later
stages is faster and typically requires less than three years \cite{KLA16, KLA16b, KLA16c, KLA18}.
Other studies conducted on ten religious radical groups by New York Police Department detectives, analysts and intelligence officials, 
show that the time spent in the at risk or in the susceptible class is approximately one and a half and three years, respectively;
the time in the more radical stages is about two years each \cite{SIL07}. Overall, radicalization times are becoming
shorter due to more rapid information sharing, especially online.

\begin{figure} [t]
\includegraphics[width=4.0in]{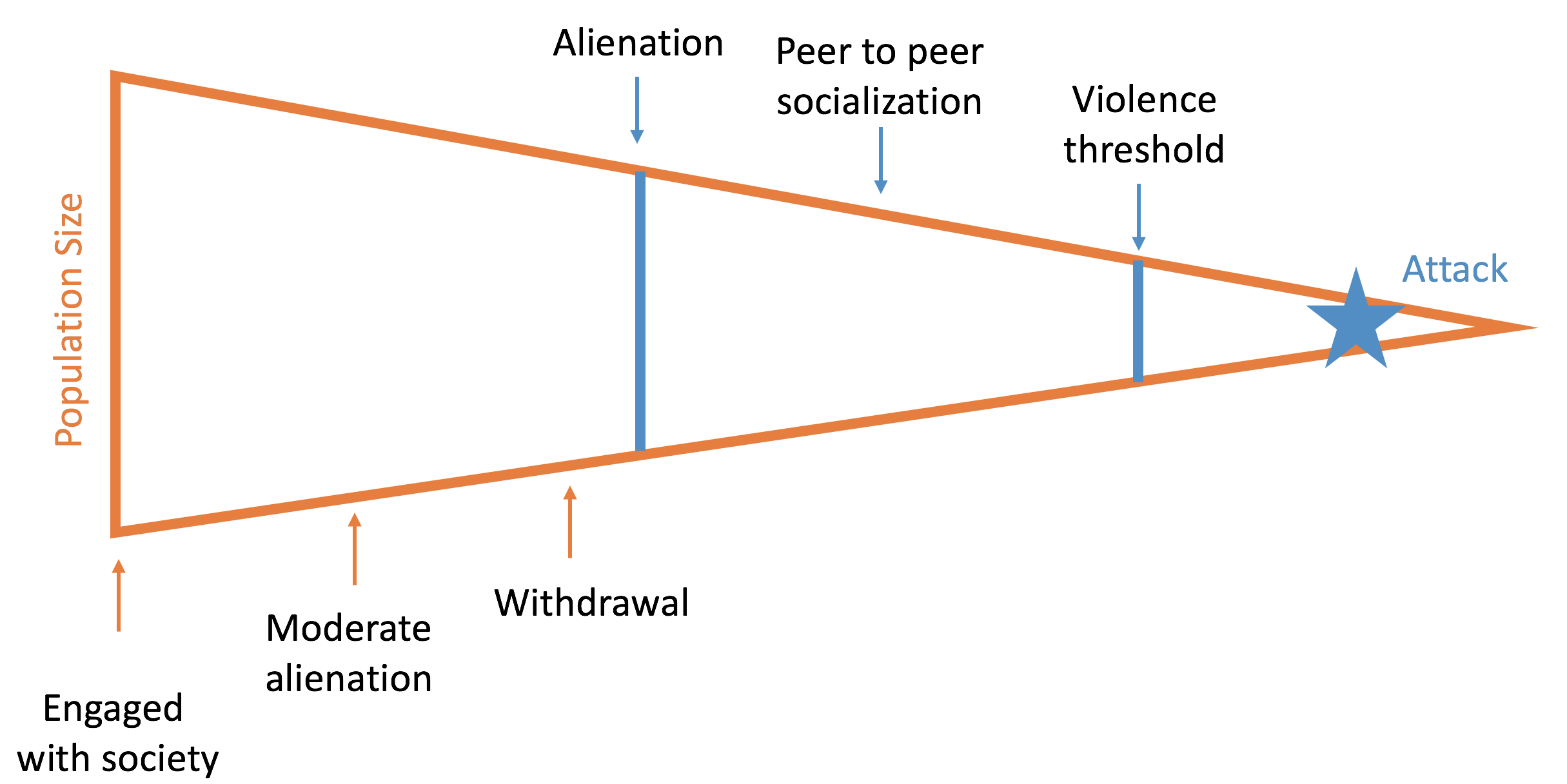}
\caption{Schematics of the radicalization process \cite{WRI06}.
Functional, engaged individuals undergo progressive phases of withdrawal until they
set themselves apart from the rest of society. Henceforth, they
cultivate a new identity, search for like-minded individuals and prepare
for violence. The process culminates in possible terrorist attacks
\cite{SIL07}. Not all individuals will progress through the entire hierarchy: 
the number of radicalizing individuals decreases as the level of extremism increases, leading to a multi-phase
horizontal funnel.}
\label{pyramid}
\end{figure}

The seed of pre-radicalization is often rooted in various unaddressed grievances and personal frustrations, such
as lack of employment and opportunity, racial and religious
discrimination and social exclusion \cite{HOR08, JAS16, HOL17, BAH24}. Individual malaise can also stem from
real or perceived socio-economic injustices, against policies that are seen
as too progressive, or conversely, because of the desire for sweeping societal change. 
Personal dissatisfaction leads to self-identification, where marginalized or self-searching individuals gradually begin constructing new identities
and routines, shifting away from old ones.  Like--minded people are actively sought and new friendships
formed \cite{DYE07}. The process is furthered by indoctrination when bonds of solidarity within these newly formed groups strengthen.
Mutual encouragement and the absence of a counter-dialectic allow for extremist views to self-reinforce and become
more engrained \cite{VON07} . Once embraced, disavowing the new ideals becomes difficult;  justification and praise of violence follow \cite{COO15}.  
A further activism phase ensues, when radicals commit to militantly spread their convictions to others until 
external events such as political or judiciary decisions, or simple serendipity, crystallize the willingness to take violent action
\cite{MEE15, GIL07}. 
A schematic of this process is shown in Fig.\,\ref{pyramid}.

\begin{figure}[t]
\includegraphics[width=4.0in]{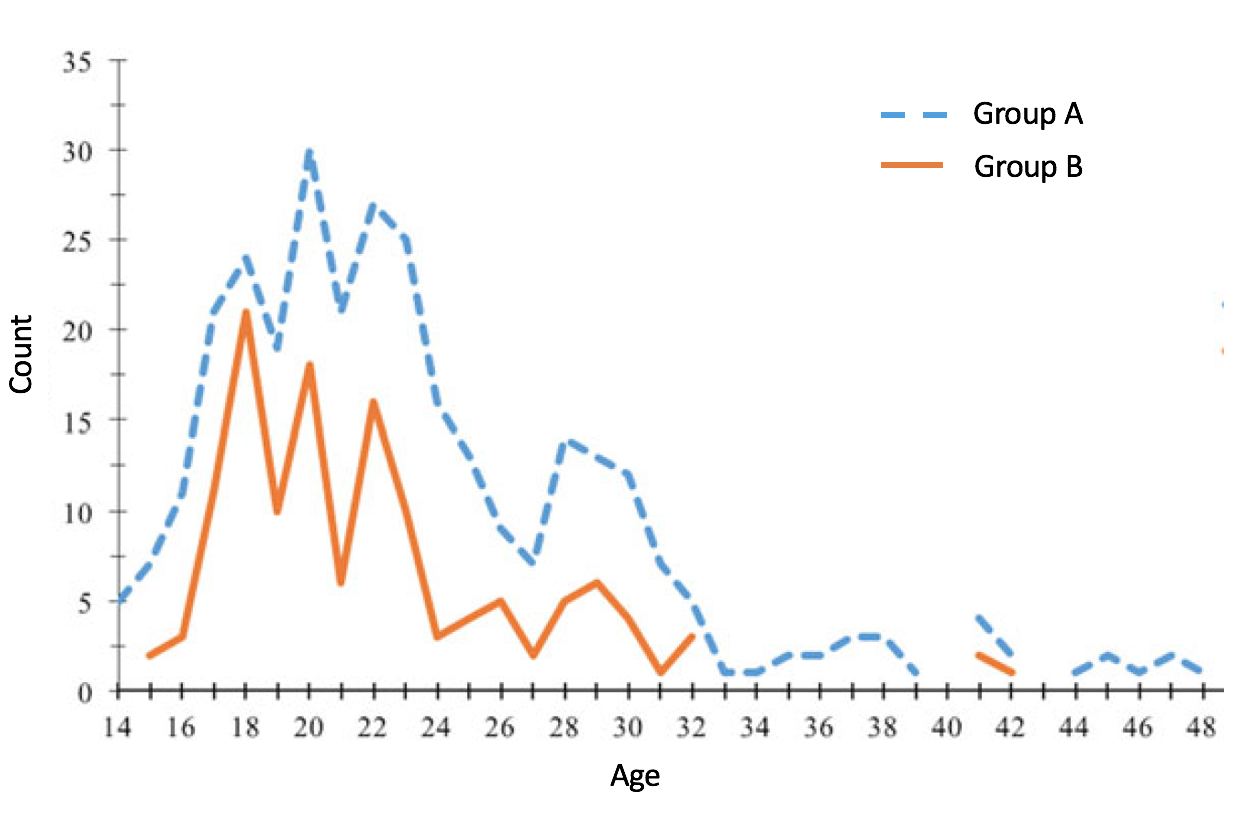}
\caption{Estimated age of interviewed subjects at the first indication of radicalization.
All were radicalized within the United States, inspired by Al-Qaeda or ISIS, and 
committed terrorist acts over a period of 16 years after September 11$^{th}$ 2001.
Group B (129 individuals)  is a subset of Group A (289 individuals) for which more detailed data was available.
Breaks indicate that no subjects of that age cohort were included in the study.  
Although individuals can be radicalized at any age, the process is more common during adolescence and early
adulthood. Taken from Ref.\,\cite{KLA18}.}
\label{agecrime}
\end{figure}

Radicalization typically occurs through networks of peers, may involve physical locations
such as prisons or mosques \cite{KAN22, RUS19, PRE24},
online tools \cite{BOT09, CON17}, and even NGOs \cite{PRE20}. 
ISIS for example is known to have extensively used social media to recruit western foreign fighters in Syria and Iraq
\cite{KLA15, BEN17}. It is easy to see why extremists exploit virtual platforms
as they offer opportunities for communication, collaboration and
persuasion, without physical contact, with anyone, at any time \cite{BEH13}. 
Extremist tendencies can arise at any age and factors that are traditionally associated with desistance from criminal or deviant
behavior, such as marriage, career and education, are not necessarily strong deterrents against ideological extremism \cite{SAG04,KIN11}.
Disenfranchised adolescents and young adults, however, are particularly vulnerable to indoctrination and radicalization,
especially when their formative years are spent without purpose, education, outlook or positive role models \cite{MEE15}.  This is
often the case for marginalized youth, religious zealots, or right-wing extremists \cite{BIZ14, KOE14}
for which radicalism may provide a sense of purpose. 
Figure\,\ref{agecrime} shows that the majority of 289 homegrown terrorists in the United States, inspired by Al-Qaeda
and ISIS, initiated their radicalization process before age 25 \cite{KLA18}.
Although in their infancy and mostly experimental, de-radicalization programs are
being developed and implemented \cite{HOR10, BAA20, ULY21, HOR20}. They 
include education, exposure to literature, sports and arts, psychological counseling, job training and informal one-on-one
conversations \cite{DEM10, CHE18, WEB20}. The goal is to prevent radicalization, disengage known violent extremists,  
and reintegration former convicted extremists in the community \cite{ELS15, FED15, SCH16, CHE16, KOE17, WEB17, WEG17}.
Intervention through social media has also been advocated: it is hoped that the interactiveness afforded by the internet can be used to
help change the perceptions of budding radicals \cite{ASH10}.  

It may be natural to ask whether parallels exist between radicalization and the process by which a propensity for crime develops.
Indeed, some studies have found similarities between becoming an extremist and joining
urban street gangs \cite{DEK11, DEK15}.  Several observations must however
be made in this respect. Terrorists, or groups of terrorists, take action much less frequently than urban criminals; 
radicals are driven by deeply held ideologies and not by opportunistic, short-term rewards;  selecting 
the appropriate times and locations, and preparing for a terrorist attack, are much more elaborate processes 
than committing street-level crimes; and finally terrorists typically seek worldwide attention in the aftermath of their acts, 
whereas burglars and other criminals prefer anonymity. Of course, the outcomes of successful terrorist and criminal
acts are also much different in terms of damages inflicted to those directly affected and to society as a whole 
\cite{TUR04, RIC06, PYS09}. Hence, although some comparisons can be made, radicalizing and becoming an urban 
criminal, should be considered distinct phenomena.

Once the radicalization process is complete, a group of like-minded extremists
may decide to turn their fanatical ideology into violent action. This requires willingness,
but also training, preparation, curating the logistics. Eventually, organizational structures emerge,
either organically or by design. The way individuals connect depends on specific needs and choices such as targets, ideology, timing, resource availability, 
governmental efficiency, public opinion  \cite{JEN18, KIL12}. One of the most important determinants of terrorist connectivity is the need to 
balance security and efficiency: on one hand, extremists must communicate and plan, 
on the other, they must protect themselves against infiltration, threats and
counter-terrorism intervention.  Governmental agencies must also decide the best course
of action to disrupt these ``dark'' networks, be it via direct repression, manipulation, or 
persuasion, and by considering several environmental variables, such
as costs, public perception, odds of success, information at hand.
The final goal may be to splinter and dissolve the terrorist network, to weaken its financial and societal support structure,
or to moderate its use of violence \cite{REE13}. Of course, 
counter-terror tactics may be counterproductive if they cause
backlash and over-reaction in the population. Thus, policies must be 
firm but not excessive or discriminatory. 

At present, two main counter-terrorism schools of thought exist.
The ``Anglo-Saxon" approach focuses on the rule of law
so that only if, and after, violent acts are committed, security agencies will intervene.
There is no curtailing of the beliefs these acts may stem from, in the name of freedom of belief
and religion. The ``continental European" approach instead attempts to prevent violence by confronting extremist ideas, 
especially the preaching of intolerance, before any attacks materialize \cite{POP02, NEU13, RIC17}. Both have merits and drawbacks,
and are deeply rooted in the history and national identity of the respective countries where
they are applied.  Finally, recent studies in political terrorism have identified ``waves'' of terrorism as 40-year generational cycles
over which radical tendencies tend to ebb and flow \cite{KAP16}. Periods of heightened activity are characterized by the
emergence of terror groups that share ideology, tactics, and vision for the future, that later subside. The wave theory of modern
terrorism was crafted by David C. Rapoport \cite{RAP02} who reviewed terrorist events from the late 1800s to the present day; 
it shares parallels with historian Arthur Schlesinger's theory of 40-year generational cycles in politics \cite{SCH86}.

The processes and theories described above are complex,  layered and highly dy\-nam\-ic, 
and involve individuals, their socio-economic environment and peer relationships, 
the larger political context. 
In the following sections, we review recent work to quantitatively 
analyze and understand them.  

\section{Radicalization in opinion dynamics models}
\label{compartment}

We begin with radicalization models framed within the context of opinion formation,
a most natural starting point since radicals can be viewed as individuals with extreme ``opinions." 
Here, the belief of an individual changes through peer-to-peer contact, the influence of media, or current events. 
Mean-field models assume that individuals with a given belief behave uniformly and define a homogeneous population; each of these
populations is assumed to interact with others influencing their views, in a reversible or irreversible manner, so that as time evolves the size of a given population
may increase or decrease.  Mathematically, the dynamics is 
described by ODEs or, in some cases, by PDEs to include spatial dependence. 
These are simplistic models, that do not take into account the much larger complexities that lure single individuals towards fanaticism, 
yet they allow for the identification of parameters or mechanisms that drive observed trends.

The work of Carlos Castillo-Chavez \cite{CAS03} is one of the first to introduce population dynamics models 
to the study of transmission of fanatical behaviors, building on 
epidemiology contact processes \cite{GOF64}. Concepts such as ``reproduction numbers'' $R_0$
used to predict when a disease becomes endemic ($R_0 > 1$) and when 
it does not ($R_0 < 1$) are adapted to the spread of radical ideologies.
In particular, a so called ``fanatic hierarchy" is introduced with the total population $T(t)$, 
divided into a subgroup $G(t)$ that has no propensity for radicalization, and three subgroups of individuals at various
stages of their commitment to the extreme ideology. The latter are $S(t)$, a group of susceptibles 
who are not yet radicalized but who are vulnerable and open to the radical ideology at hand; the population $E(t)$, 
individuals who have recently turned into 
fanatics; and finally $F(t)$ individuals who have completely espoused radical views.
The various subpopulations are related by $T = G + S + E + F$, and 
the model is written as

\begin{eqnarray}
\label{CC1}
\frac{dG}{dt} &=& \Lambda - \beta_1 G \frac{C}{T} + \gamma_1 S + \gamma_2 E + \gamma_3 F - \mu G, \\
\label{CC2}
\frac{dS}{dt} & =& \beta_1 G \frac{C}{T} - \beta_2 S \frac{E + F}{C} - \gamma_1 S - \mu S, \\
\label{CC3}
\frac{dE}{dt} & =& \beta_2 S \frac{E + F}{C} - \beta_3 E \frac{F}{C} - \gamma_2 E - \mu E, \\
\label{CC4}
\frac{dF}{dt} & =& \beta_3 E \frac{F}{C}  - \gamma_3 F - \mu F,
\end{eqnarray}

\begin{figure}[t]
\begin{center}
\includegraphics[width=3.0in]{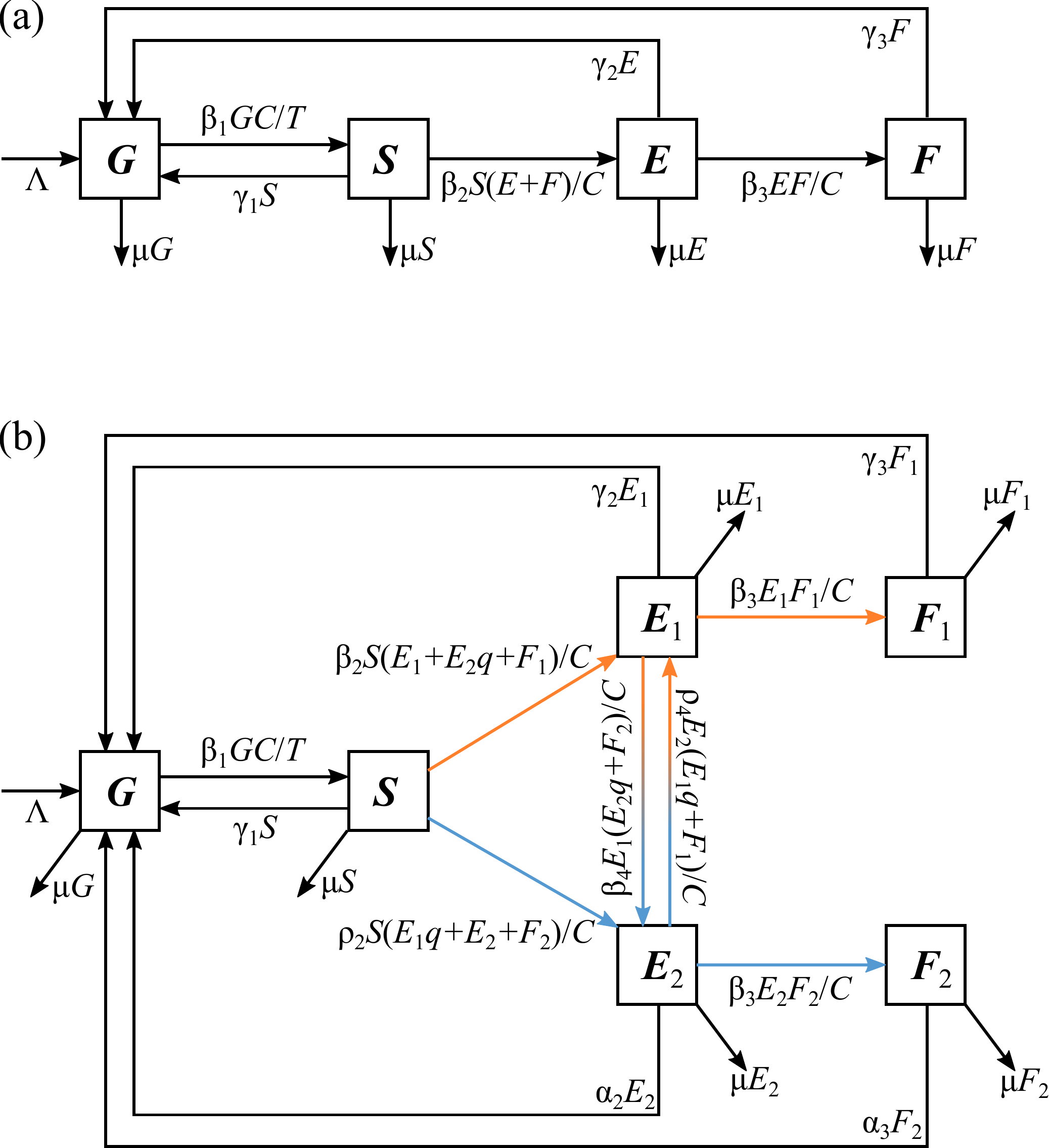}
\end{center}
\caption{Schematics of the radicalization process according to representative compartment models \cite{CAS03, CAM13}.
In panel (a) individuals may progress from a general, non-radical state
$G$ towards a hierarchy of susceptibles $S$, recent adherents $E$ and full fanatics $F$. Birth and death are also included.
In panel (b) two radical groups $\{E_1, F_1\}$ and $\{E_2, F_2\}$ can originate from the susceptible cohort $S$. The two ideologies 
interact with adherents moving between recent convert groups $E_1, E_2$. The parameter $q < 1$ indicates how effective
groups are at cross-recruitment. The transition $E_2 \to E_1$ is modulated by the full radical cohort $F_1$ and by a percentage
of recent converts $q E_1$; the $S \to E_1$ transition is modulated by $E_1 + F_1$
and by some members of the opposite group $q E_2$ who proselytize in favor of their counterparts.
Similar arguments hold for $E_1 \to E_2$ and $S \to E_2$. In both panels, $T$ is the total population
and $C = T - G$.}
\label{CasCam}
\end{figure} 

\noindent
where $C(t) = T(t) - G(t)$ represents the collection of individuals who are susceptible, partially or fully 
radicalized. The system includes a birth term $\Lambda$ into the non-susceptible population $G$
and a death rate $\mu$ for all subpopulations, so that $dT/dt = \Lambda - \mu T$, yielding $T(t \to \infty) = \Lambda/\mu$. All other terms 
in Eqs.\,\eqref{CC1}--\eqref{CC4} are associated with specific transitions between the various subpopulations. 
For example, individuals of the non-susceptible population $G$ may become receptive to radical ideologies if exposed
to them; the push of the radical message is modeled via the population ratio $C/T < 1$ modulated by the rate $\beta_1$
so that the net flow $G \to S$ is expressed by $\beta_1 G \,C/T$.  
Similarly, the flow from the susceptible state into the first stages of fanaticism $S \to E$
depends on the indoctrination and on the example set by the more radical individuals and is given by $(E+F)/C$ modulated
by the rate $\beta_2$. The transition from the initial radical state to the fully committed one 
$E \to F$ is instead driven solely by the radical cohort via $F/C$ and modulated by the
rate $\beta_3$. The model is summarized in Fig.\,\ref{CasCam}\,(a).

Eqs.\,\eqref{CC1}--\eqref{CC4} define a hierarchy in the sense that transitions through the 
$G \to S \to E \to F$ stages are propelled by more radical populations influencing the lesser ones.
The three rates $\gamma_i$, for $i=1,2,3$, represent de-radicalization from the 
susceptible, recent convert, and fully fanatical subgroups, respectively. Note that there are no 
reverse intermediate transitions: radical individuals $F$ for example
may return to the non-susceptible population $G$ but do not transition back to their first stage of radicalization $E$.

Despite its simplicity, the model in Eqs.\,\eqref{CC1}--\eqref{CC4} offers useful insight on societal
outcomes, in terms of attractor points and thresholds.  Of course the most important question to ask 
is under which conditions does a finite fanatic population $F$ arise. Analysis of Eqs.\,\eqref{CC1}--\eqref{CC4}
shows that no level of radicalization will be sustained $(S=E=F=0)$ for
 $\gamma_1 > \beta_1$, implying that one way to avoid radical discourse in a society 
is to hamper the onset of the radical process at the early $G \to S$ stage, when individuals become
susceptible to extremism. Depending on other parameter combinations, attractor points with no fanatic populations $(S^* \neq 0, E=F=0)$ or
$(S^* \neq 0, E^* \neq 0, F=0)$, and with fanatic populations $(S^* \neq 0, E^* \neq 0, F^* \neq 0)$ can be identified.
Equally important are the initial conditions and time scales involved: for example, even
if the steady state is predicted to yield $F=0$, an initially small cohort of extremists can 
successfully invade the population and lead to a large fanatic population over intermediate times,
before it begins to decay.  

Numerous studies have built upon Eqs.\,\eqref{CC1}--\eqref{CC4}.  
The possibility of two competing radical groups emanating from the same general, non-susceptible
population $G$ has been modeled by including two recruitment rates $q \beta_1$, and 
$(1 - q) \beta_1$ feeding into two distinct susceptible groups \cite{CAM13}. 
Each of these two cores follows the same radicalization process illustrated in Eqs.\,\eqref{CC1}--\eqref{CC4} 
without interacting. More interestingly, the two subgroups may also 
experience cross-contact across the $S \to E \to F$ hierarchy as illustrated in a different variant of the model shown
in Fig.\,\ref{CasCam}\,(b).
Interactions between the two radical branches represent recruitment competition or inadequate retention efforts
so that fanatics may draw new adherents from converts to the opposite ideology. 
Trade-offs between recruitment and retention are addressed, and parameter regimes whereby one of the two ideologies will persist  
and the other will become extinct, are identified. In particular, it is shown that the two groups cannot both coexist
at steady state without cross interactions: competitive recruitment is necessary for the emergence of
two finite fanatic populations, $F_1, F_2$.  Successive studies include simplified models 
that focus on the recruitment process outlined in \cite{CAS03, CAM13}, that include 
de-radicalization treatments of extremists \cite{SAN18, MCC18, SAN19}, that parametrize
government intervention \cite{NAT18}, or that include coercion, kidnapping
and other means of involuntary recruitment into the terrorist organization \cite{GAM20, SMA22}.

Compartment models akin to \cite{CAS03, CAM13} have also been applied to actual 
extremist movements, for example to study the influence of separatist groups in
the Basque country of Spain \cite{SAN08, EHR13}. Subpopulations were created and 
transition parameters were estimated using demographic and electoral data;
immigration and emigration terms were included as well. The ideological evolution 
of society was thus analyzed over a 35-year period, including projections into the future. 
Similarly, Eqs.\,\eqref{CC1}--\eqref{CC4} were adapted to the study of radicalization of extreme
right-wing movements in Germany \cite{DEU15} where 
Likert scale surveys repeated biannually from 2002 onwards \cite{LIK32,ALL07} 
were used for calibration. In a separate study, data on ten violent extremist online forums was
collected over four years to develop a visitor-engagement model \cite{CEB13}. It was found that
users undergo several layers of involvement until they become recruiters or real-life terrorist operatives. 
The authors discuss strategies to control the proliferation of online groups,
and find that aiming for the dissolution of a large number of extremist
cyber-communities is not very effective as enforcement costs increase but
gains are limited. Targeted, occasional intervention towards large groups is a
preferable policy, as it will induce other, minor, forums to self-regulate. 

Alternate compartment models define subpopulations along cultural lines,
so that responses to given societal issues may fuel radicalization in more sensitive
subgroups \cite{GAL16}. Here, the core subgroup represents the mainstream;
two sensitive subgroups include first to third generation immigrants whose ways of life may be partially incompatible
with the core. Once conflict arises, core agents are assumed to be inflexible; the sensitive subgroups
may choose to adjust to mainstream beliefs and attitudes, or
to oppose them. Three categories thus arise: core-inflexible $\sigma_{I}$; sensitive-peaceful, $\sigma_{P}(t)$, 
and sensitive-opponent $\sigma_{O}(t)$.  The latter subgroup is assumed to be the source of radical activity. 
Sensitive individuals can change their status depending on interactions among themselves as well as with the core,
according to the following dynamics

\begin{eqnarray}
\label{javone}
\frac{\dd \sigma_{P}}{\dd t} = \alpha \sigma_{I} \, \sigma_{O} - \beta \sigma_{O} \sigma_{P}, \quad
\frac{\dd \sigma_{O}}{\dd t} = -\frac{\dd \sigma_{P}}{\dd t},
\end{eqnarray}

\noindent
where the $\sigma_{i}$ values, $i = \{I, P, O \}$ represent subpopulation fractions so that
$\sigma_I + \sigma_{P}(t) + \sigma_{O}(t) = 1$.  Since the core is inflexible, 
the value of $\sigma_I$ is fixed. In Eqs.\,\eqref{javone}
opponents recruit peaceful individuals but this process is hindered by the 
core that tends to steer opponents towards the peaceful state.  
The authors investigate the conditions under which a small initial minority of opponents 
persists, leading to long term radicalization. Of the two steady states
that arise, only one has a finite opponent population. 
In particular, for a large enough core $\sigma_I > \beta/(\alpha + \beta)$, 
opponents $\sigma_{O}(\infty) \to 0$, precluding radicalization.
Although the model in Eqs.\,\eqref{javone} is limited to exchanges between segments of society and does not include
institutional roles, its results show that radicalization may be reduced by core citizens ($\sigma_I$) 
engaging with sensitive agents ($\sigma_{O}, \sigma_{P}$).
The inflexible core $\sigma_I$ places this work among the many opinion dynamics models
where a ``zealot'' fraction of the population holds a fixed view that can never be changed,
and where final outcomes depend on its magnitude \cite{MAR12, VER14}.

Other compartment models have included possible catalysts for radicalization 
such as unpopular political decisions or foreign interventions.
These may generate hatred that spreads within a population, leading to outbreaks of violence. 
For example, in \cite{NIZ14}, five possible states describe societal attitudes towards a given issue or decision; 
the population can be ignorant but sensitive to the topic at hand,  and once aware, 
immune, upset, or violent.  All five groups influence each other and contagion unfolds 
from an initial condition of a small cohort of those upset by the news interacting with those who
are susceptible or immune to it.
After being upset, and then violent, individuals may accept
the \textit{status quo} and become relaxed. Particular attention is given to the growth of the
violent cohort whose ranks increase as the upset population interacts with the rest.
Finally, compartment models have also been used
to study the internal dynamics of terrorist cells, with 
interacting groups of leaders and foot-soldiers; counter-terror measures targeting
both groups are also included in the form of distinct death rates \cite{GUT09}.

Following a slightly different approach, the emergence of two antagonistic, radicalized groups was modeled in \cite{SHO17}. 
Society is here assumed to host two competing religious, ethnic or political groups, each of them harboring a moderate and a 
radical faction that interact with each other. Included in the model is the possibility of one group violently attacking the other.  
Intra-group transitions within a sect (from moderate to radical, and vice-versa) are assumed to occur either spontaneously or through
indoctrination, but also depend on the actions and characteristics taken by the other sect. This is done by including a
sensitivity parameter that indicates how strongly individuals will react to external 
attacks or propaganda.   
The two $A$ and $B$ sects are divided into time dependent $r_A (t), r_B (t)$ radical 
and $n_B(t), n_B(t)$ moderate factions.  Since the total population per sect is assumed to be constant, and
$n_A(t) + r_A(t) = N_A$ and $n_B(t) + r_B(t) = N_B$, the time evolution of each sect is completely determined
from the dynamics of the radical component. For sect $A$ thus

\begin{eqnarray}
\label{sects2}
\frac{\partial r_A}{\partial t} =n_A \left[ \lambda_A s_A + p_A s_A \frac{r_A}{N_A} - \mu_A f (s_A) r_A \right], 
\end{eqnarray}

\noindent
while sect $B$ follows the same dynamics with $A \to B$ and vice-versa.
The first term on the right-hand side of Eq.\,\eqref{sects2} represents spontaneous radicalization described
by the intrinsic rate $\lambda_A$, modulated by the sensitivity parameter $s_A$. The second term represents
radicalization in response to indoctrination from the $r_A/N_A$  fraction of active radicals at rate $p_A$, and
similarly modulated by $s_A$. De-radicalization is assumed to occur at an intrinsic rate $\mu_A$ modulated
by the sensitivity-dependent function $ f(s_A)$. The latter
decreases with $s_A$ so that a highly sensitive population is more unlikely to de-radicalize.  
Finally, the sensitivity parameter $s_A$ quantifies how strongly sect $A$ reacts to attacks originating from sect $B$. 
To model couplings between the two sects we first assume that sect $B$ attacks sect $A$ at rate $k_B r_B$, where $k_B$ is bounded
by a maximum threshold $\omega_B$ due to various limitations,  so that $k_B < \omega_B$. 
We then set $s_A = ( k_B \, r_B ) / ( \omega_A N_A )$.   
This relationship implies that sensitivity depends on the ratio between the actual attack rate, 
via the term $k_B r_B $, and the hypothetical maximum retaliation rate sect $A$
could exert on sect $B$, represented by $\omega_A N_A$.
In this context, $s_A$ serves as a measure of the relative aggression capabilities of the two groups.
The parameters $k_B$, $\omega_A$ enter the model only through the sensitivity term $s_A$; 
similarly for $k_A$, $\omega_B$ appears in the model only through $s_B$. 
In scenarios where the sensitivity of the two factions are independent of each other’s actions, both $s_A$ and $s_B$ 
can be treated as constants, and the dynamics of the two groups become effectively decoupled. 

The dynamics that unfolds from Eq.\,\eqref{sects2} is rich: 
steady states, stability and bifurcations are determined under several scenarios of sect size and interactiveness.
A game theoretic framework \cite{STA11, JOV88, LAS06a, LAS06b} is further added, assuming that radical factions may tune ``strategic'' parameters,
such as intensity of propaganda $\lambda_{A,B}$ or frequency of attacks $k_{A,B}$ to optimize given utility functions 
aimed at increasing rank numbers while decreasing rival attacks.  Resources are assumed to be limited so that choices must be made between 
proselytizing and attacking; a sect can also readjust its choices in response to the actions
taken by the other.   The main finding of this work is that unless very high rates of violence are employed, the existence
of relatively small groups of radicals cannot be sustained: eventually, they become less extreme and only the moderate faction will persist.
The perspective that emerges is that to survive, radical groups must employ greater violence in their early days, when they are still numerically small, and may
transition towards less violent methods, such as indoctrination of moderates, later on, as they mature.  

Other compartment models consider the coevolving dynamics of terrorism and illicit drug use \cite{AKA22a, AKA22b, AKA23},  
terrorism and corruption \cite{BAB23},  the rise of insurgents \cite{KOL20}, 
and the spread of extremism in a society where moderates
tend to avoid confrontation \cite{GAL22}. Finally, system dynamics approaches to understand how transnational terrorist groups 
recruit new members, train and sustain them, and execute terrorist attacks have also been introduced, 
in particular to study the activities of Al-Qaeda and responses from the 
United States \cite{CHA07, CAR14}. Non-linear behaviors are analyzed via 
so called stocks and flows, feedback loops and time delays, allowing for several scenarios
to be explored through graphical user interfaces \cite{EGG02, OGA03}. 

\section{Radicalization and age structure}
\label{agestructure}
The mathematical models described above provide many useful insights, however none 
include age-sensitive responses to propaganda, emulation of peers and societal pressure. As described in the introduction, 
radicalization is highly age-dependent, and age stratified models may help better understand
the establishment and evolution of radical groups over time. 
A compartment model of increasingly fanatic stages coupled to age-differentiated interactions
was introduced in \cite{CHU17}. For simplicity, only three cohorts are considered in this work: 
non-radical $i=0$, activist $i=1$ and
radicals $i=2$ described by the densities $\rho_i (t, a)$  
of age $a$ at time $t$ for $i = \{0,1,2 \}$.
Transitions between successive pools 
are mediated by the activation rate $A(a; \rho_1)$ (from $i=0$ to $i=1$) 
and the radicalization rate $R (a; \rho_2)$ (from $i=1$ to $i=2$).
Both rates depend on the age of the source population
as well as on the structure of the sink population. 
A de-activation rate $C_D$ (from $i=1$ to $i=0$) and a pacifying rate $C_{\textrm P}$ (from $i=2$ to $i=1$)
are also included and assumed to be age-independent for simplicity. The 
resulting age-structured model is of the McKendrick-von Foerster type 
\cite{MCK25, LES45, LES48, TRU65, KEY97, CHO16} and is written as

\begin{eqnarray}
 \label{EQ:RHO0} 
   \frac{\partial \rho_0}{\partial t} + \frac{\partial \rho_0 }{\partial a} &=& 
   - A (a; \rho_1) \rho_0 + C_{\textrm D} \rho_1,  \\
   \frac{\partial \rho_1} {\partial t} + \frac{\partial \rho_1 }{\partial a} &=& A (a; \rho_1) \rho_0 
    - \left[ C_{\textrm D} + R(a; \rho_2) \right] \rho_1 +
    \label{EQ:RHO1}  
  C_{\textrm P} \rho_2.
\\
  \frac{\partial \rho_2 }{\partial t} + \frac{\partial \rho_2 }{\partial a} &=&  
   \label{EQ:RHO2} 
   R (a; \rho_2) \rho_1  - C_{\textrm P} \rho_2. \\
   \nonumber
\end{eqnarray}

\noindent
The left-hand side is the total time derivative $d/dt  = \partial / \partial t +  \left( \partial a / \partial t  \right) \partial / \partial a$;
provided age and time are measured in the same units $\partial a / \partial t  = 1$.
The transition rates that appear above are expressed as

\begin{eqnarray}
A (a; \rho_1) &=& C_{\textrm A} \int_{a_0}^{a_1}   
   {\mathcal K} (a,a'; \alpha_A, \sigma_A) \, 
   \rho_1 (t, a^\prime) \dd a^\prime,
   \label{EQ:ACTIVATION_RATE} \\
   R (a; \rho_2) &=&  C_{\textrm R} \int_{a_0}^{a_1} 
   {\mathcal K} (a,a'; \alpha_R, \sigma_R) \, 
   \rho_2 (t, a^\prime) \dd a^\prime,
   \label{EQ:RADICALIZATION_RATE}
\end{eqnarray}

\noindent
where the interaction kernels ${\mathcal K}(a,a', \alpha_j, \sigma_j)$ 
couple populations of different ages

\begin{eqnarray}
\label{kernels}
      {\mathcal K}(a,a'; \alpha_j, \sigma_j) = \left[\displaystyle{\int_{-\bar{a}}^{\bar{a}} e^ {- s^2 / \sigma_j^2}\, \dd s} \right]^{-1}
    \displaystyle{\exp {\left[ - \frac{(\alpha_j - a)^2+
            (a - a^\prime)^2}{2 \sigma_j^2} \right]}},
\end{eqnarray}

\begin{figure}[t]
\begin{center}
\includegraphics[width=3.0in]{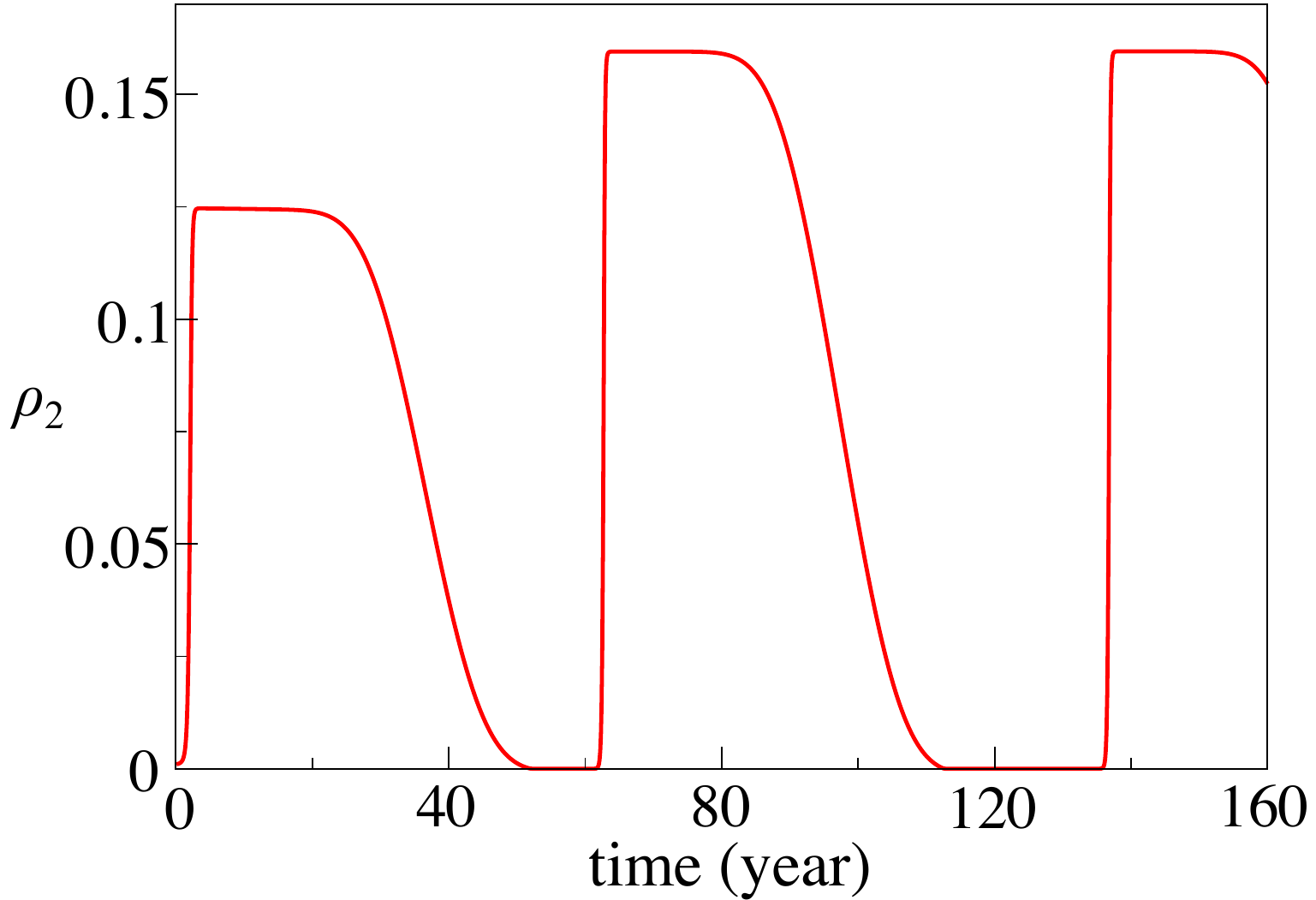}
\end{center}
\caption{Time dependence of the radical population $\rho_2(t) = \int_{a_0}^{a_1} \rho_2(t,a) da$
arising from Eqs.\,\eqref{EQ:RHO0}--\eqref{EQ:BIRTH1} under irreversible
radicalization $C_{\textrm P} = 0$. Radical populations oscillate defining 
40-year cycles reminiscent of Rapoport's wave theory of modern terrorism \cite{RAP02}.
 Other parameters are $a_0 = 5$ years, $a_1 = 55$ years,  $C_{\textrm R} = 50$, $C_{\textrm A}= 12$, $C_{\textrm D} = 5$,
$\alpha_{\textrm A} = \alpha_{\textrm R} = 20$ years, $\sigma_{\textrm A} = \sigma_{\textrm R} = 10$ years; 
initial conditions are $(\rho_0 (a,0), \rho_1 (a,0), \rho_2 (a,0))=(0.989, 0.01, 0.001)$. 
Under these parameters, radicalization is aggressive, $C_{\textrm R} > C_{\textrm A} > C_{\textrm D}$, and leads to a buildup
of extremists $\rho_2(a,t)$ that depletes the activist pool $\rho_1(t,a)$ until no further recruitment 
of the general population $\rho_0(t,a)$ is possible. The radical
cohort eventually wanes due to aging; the cycle restarts with the next generation.}
\label{oscillate1}
\end{figure}

\begin{figure}[t]
\begin{center}
\includegraphics[width=2.5in]{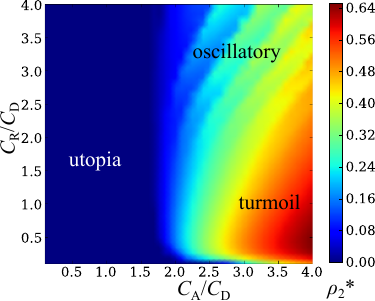}
\end{center}
\caption{Long-time radical populations of any age as a function of $C_{\textrm A}/C_{\textrm D}$ and $C_{\textrm R}/C_{\textrm D}$ 
and under irreversible radicalization $C_{\textrm P} = 0$, 
for the fanatic-stage model of age-differentiated interactions in \cite{CHU17}.
In addition to utopia (no radicals present) and turmoil (radicals persist) 
an oscillatory regime arises at steady state for large values of $C_{\textrm R}/C_{\textrm D}$
with $C_{\textrm R} > C_{\textrm A}$. 
Here, periods of turmoil yield to dormant phases, defining 40-year cycles in accordance with the
wave theory of modern terrorism \cite{RAP02} and as shown in Fig.\,\ref{oscillate1}. Other parameters 
are $a_0= 5$ years, $a_1 = 55$ years, $C_{\textrm D}= 5$, $\alpha_{\textrm A} = \alpha_{\textrm R} =  20$ years, 
$\sigma_{\textrm A} =  \sigma_{\textrm R} = 10$ years.}
\label{oscillate}
\end{figure}

\noindent
for $j = {\textrm A, R}$. Although many choices are possible, as written, the interaction kernels 
indicate that individuals are most
susceptible to activation and radicalization at target ages $a= \alpha_ j$ for $j={\textrm A,R}$
and through ``peer to peer" interactions with individuals of similar age but more radicalized; typical values
are 20 to 30 years old. The ranges $\sigma_j$ represent the effective age overlap between
populations of different ages. The overall span of radical activity is assumed to occur within the $\left[ a_0, a_1 \right]$ window,
so that individuals of age $a < a_0$ are too young to influence or be
influenced by their peers, while those with age $a > a_1$
may be too old or entrenched for change.  The term $\bar{a} \equiv a_1 - a_0$ in the prefactor of Eq.\,\eqref{kernels}
guarantees that upon integration over $[a_0, a_1]$ the kernels are independent of
$\sigma_{j}$. Finally, boundary conditions are set as

\begin{eqnarray}
  \rho_1 (t, a_0) & = & \rho_2 (t, a_0) = 0,  \qquad
   \rho_0 (t, a_0) = \sum_{i=0,1,2} \rho_i (t, a_1) \label{EQ:BIRTH1}.
\end{eqnarray}

\noindent
The model defined by Eqs.\,\eqref{EQ:RHO0}--\eqref{EQ:BIRTH1} was first analyzed by 
integrating over the age interval $[a_0 ,a_1]$, yielding a compartment model for which steady states
were determined. Later the full age-dependent model was analyzed to draw comparisons.
For both versions of the model, age-independent and age-dependent, 
three steady states arise: utopia (where society is solely comprised of non-radicals, $\rho_0^* \neq 0$, $\rho^*_1 = \rho^*_2 = 0$),
a dormant state (with finite populations of non-radicals and activists $\rho_0^* \neq 0$,  $\rho^*_1 \neq 0,  \rho^*_2 = 0$) 
and turmoil (where radicals are also present $\rho_0^*\neq 0$, $\rho^*_1 \neq 0,  \rho^*_2 \neq 0$) . 
However, the respective basins of attraction
and the associated populations differed greatly in the two formulations. For example, 
including age structure enhances radicalization in certain parameter regimes, accelerating the transition through the $i=0, 1, 2$ hierarchy. 
In other cases, precocious radicalization drains the activist pool and hinders further radicalization. The latter
scenario arises when the radical ideology spreads too quickly among a few fanatics who become isolated from the rest of society, 
and therefore unable to recruit more adherents through peer pressure at the intermediate, activist level.

Most notably, age-independent radicalization models do not display limit cycles  whereas, the model
in Eqs.\,\eqref{EQ:RHO0}--\eqref{EQ:BIRTH1} allows for cyclic behavior to arise. Alternating
waves of more or less radicalized individuals are observed over the course of several generations  as shown in Fig.\,\ref{oscillate1}.
This result aligns with political science paradigms first presented by Rapoport and discussed in the introduction
whereby extremist tendencies rise and fall over time like waves or ripples \cite{RAP02, SED07, MCA11, PAR16}.
Eqs.\,\eqref{EQ:RHO0}--\eqref{EQ:BIRTH1} provide possible mechanisms to explain wave-like behavior; 
cyclic solutions typically arise when radicalization is aggressive ($C_{\textrm R} > C_{\textrm D}$ and
$C_{\textrm R} > C_{\textrm A}$) and irreversible ($C_{\textrm P} = 0$).  Under these conditions 
the intermediate activist population $\rho_1(a,t)$, which serves as a bridge between
non-radicals and radicals, is depleted, preventing non-radicals from even being exposed to fanatic ideals
and entering the radicalization cycle in the first place. This scenario represents radicals
becoming so extreme that they eventually alienate moderates and the general population,
jeopardizing their own survival. Without fresh recruits, the extremist movement fades away due to aging. 
Realistic parameters result in a period of sustained radical populations of 40 years, in agreement with 
Rapoport's theory according to which extremists radicalize a generation of individuals, and when their influence fades due to aging, 
the cycle of terrorism comes to an end \cite{RAS09}.
Simulation outcomes in parameter space are displayed in Fig.\,\ref{oscillate}, showing long time
radical populations of any age as a function of $C_{\textrm A}/C_{\textrm D}$ and $C_{\textrm R}/C_{\textrm D}$ 
under irreversible radicalization $C_{\textrm P} = 0$. Small $C_{\textrm A}/C_{\textrm D}$ leads to
utopia; turmoil prevails at small $C_{\textrm R}/C_{\textrm D}$ and large $C_{\textrm A}/C_{\textrm D}$. 
For large $C_{\textrm R}/C_{\textrm D}$ and intermediate $C_{\textrm A}/C_{\textrm D}$
society oscillates between utopia and turmoil, defining 40-year cycles.

We conclude by noting that most sociological theories model radicalization as a linear process,
and assume that the adoption of extreme beliefs is a necessary condition for the execution of violent acts. 
Profiles of actual terrorists show that this is not always the case \cite{NEU13, BOR12}, and 
that  individuals who hold radical views may opt to peacefully manifest their discontent.  
Furthermore, social movement theory suggests that terrorism is a small part of a much 
wider array of potential actions arising from protest or counterculture movements \cite{MAC01,WIK03}. The radicalization pathway 
in Fig.\,\ref{pyramid} can thus branch out towards multiple end points, of which only a subset includes violence.

\section{Lattice and network models}
\label{networks}

\begin{figure}[t]
\hspace{-0.5cm}
\includegraphics[width=4.5in]{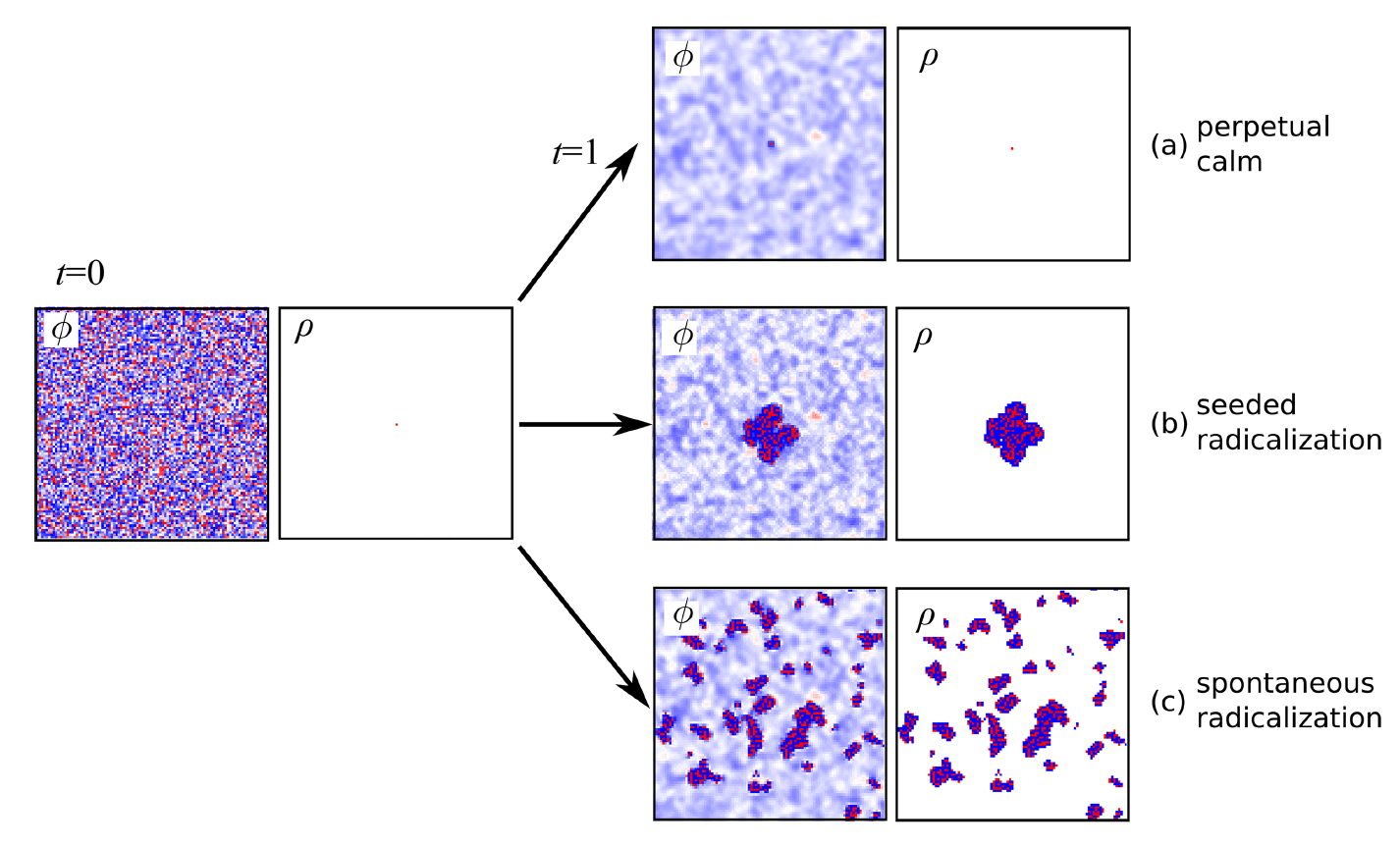}
\caption{Radicalization processes on a square lattice. At $t=0$
opinions $\phi$ are heterogeneously distributed
and one radical element $\rho \neq 0$ is present at the center of the lattice.
Parameter choices lead to three different possible scenarios.
In panel (a) under high tolerance for different opinions, views are heterogeneous
but no radicals emerge; this is the perpetual calm scenario. In panel
(b) under intermediate tolerance the initial radical is able to nucleate
a cohort of other extremists that at $t \to \infty$ will cover the entire lattice;  this is
a seeded nucleation event. Finally, in panel (c) individuals are intolerant to 
different views and clusters of radicals emerge throughout the lattice; this 
is the spontaneous radicalization case.  Under perpetual calm, the radical seed is unable to
radicalize anyone else; under seeded radicalization, a radical
population emerges from the initial seed through
nearest-neighbor interactions. Under spontaneous radicalization, individuals
radicalize even in the absence of direct contact with other
radicals. Taken from Ref.\,\cite{CHU17b}.}
\label{seed}
\end{figure}

About fifty years ago, experiments on collective decision-making revealed
an unexpected sociological phenomenon, termed group polarization. 
When faced with diverse opinion or judgement possibilities, 
a medium-sized group of interacting individuals did not settle on a moderate, intermediate view.
Rather, interactions lead the group to find consensus on an extreme position, the one 
supported by its most polarized members; such outcome was more pronounced if the topic was complex and nuanced \cite{MOS69, MOS72}.  
Several theories were offered to explain this discovery
including diffusion of responsibility, persuasion, familiarization, and cultural values  \cite{DIO70, FRA71, DOI74}.
Inspired by this work, Serge Galam and Serge Moscovici introduced one of the first lattice type models
applied to social decision-making with the aim of describing group polarization \cite{GAL91}. 
The model is reminiscent of a classical Ising model for spontaneous magnetization,
where entropy, couplings, temperature are given sociological interpretations. Individuals are
assigned a spin-type opinion value of $\pm 1$ and interact on a square lattice in the presence of 
site-dependent social fields $S_i$ that exert a non-uniform pressure. Neighboring sites
may thus intrinsically favor different spin values, but the spin-spin couplings may drive them towards a consensus. 
Individuals can also interact with other groups, beyond their nearest neighbors. 
The size of these groups is fixed, and interactions are included in the social field.
The authors find that smaller interacting groups, with less vigorous spin-spin couplings, or that allow for 
less dissent (the low temperature scenario) tend to polarize or ``break the symmetry'', whereas larger or
more tolerant groups (the high temperature scenario) yield a more moderate consensus. 
This work was one of the first to study evolving opinions on a square lattice and 
showed the importance of the underlying social connectivity 
(couplings and interaction group size) in determining the spread and persistence of 
extreme views. 

Radical behavior often germinates within small fringe groups, or is championed
by single individuals, who are able to 
influence large segments of society depending on persuasion and delivery methods, 
but also on their connectivity.  Thus, the inclusion of structured interactions between people or communities, 
as inspired by Galam and Moscovici's work,
is a natural way to advance the study of radicalization beyond compartment models \cite{FAR07, CIO11, CHU17b}. 
Typically, agents on a lattice interact with their neighbors and
go through a progression of stages of increasing extremism. 
For example, opinions may define a
continuous variable $\phi$ between $\pm 1$,  allowing for neutral ($\phi \sim 0$) and polarized ($\phi = \pm 1$) views to coexist \cite{CHU17b}.
An individual with an extreme opinion is not necessarily a radical, since he or she may harbor
a polarizing view while still being tolerant of the opinion of others.  In this context, radicalization
depends on how one responds to views that diverge from one's own. 
Mathematically, each individual can be modeled as being subject to a tension
that depends on the difference between its opinion and that of its neighbors; 
reactions to diverse views are modulated by a tolerance level so that 
large enough tension and low tolerance lead to radicalization.
Simulation outcomes of such a scenario are shown in Fig.\,\ref{seed}. Under high tolerance, a non-polarizing
consensus is reached and no radicals exist, akin
to the emergence of an average opinion in compartment models. This is the perpetual calm
scenario of Fig.\,\ref{seed}\,(a). When individuals are intolerant to opinion heterogeneity,
radical individuals $\rho$ may spontaneously emerge, as depicted in the spontaneous radicalization
panel in Fig.\,\ref{seed}\,(c). Under intermediate levels of tolerance,
only if the lattice has been originally nucleated with a radical element will further radicals
appear, underlying the importance of initial conditions. 
This is the seeded radicalization case of Fig.\,\ref{seed}\,(b).

Although not without controversy, it is believed that typical social interactions 
such as the World Wide Web define scale-free networks,
where the probability distribution $f(k)$ for a node to have $k$
links to others follows a power law $f(k) \sim k^{- \gamma}$ \cite{ALB99, BAR99, ALB02, NEW03, CLA07}. 
Here, $\gamma$ measures how well-connected the network is; typical ranges
are $2 \leq \gamma \leq 3$. In a scale-free network a few nodes
have many connections, the majority have only a few, and there is no sharp boundary between these
two possibilities. 
Scale-free networks have been used to study the spread of ideologies, 
for example through voter models \cite{CLI73, HOL75}. Here each node represents an individual
that carries a 0 or 1 opinion that adjusts depending on the opinion of the nodes it is connected to.
These interactions often lead to consensus, where a single opinion
permeates the entire graph. To account for scenarios where conflict arises, such as in political
debates, ``zealots'' are introduced similarly as in Eqs.\,\eqref{javone}.
These stubborn players never change their attitudes and allow for distinct opinions to persist \cite{MOB03, MOB07, KUH13}.  Other
types of networks to study how cascades of extreme views can percolate throughout  
society have also been considered \cite{LIP06, BEN09, RAM15}. In particular,
small-world network models of interacting, stubborn nodes have also been shown to 
allow for radical views to persist; demographic surveys conducted 
on religious perspectives and on the state of the economy in different countries 
are consistent with model outcomes \cite{RAM15}.

The radicalization progression first proposed by \cite{CAS03}
and modeled via Eqs.\,\eqref{CC1}--\eqref{CC4} was adapted to two scale-free networks with $\gamma = 3$.
The two cases considered are a hierarchical network with directed
connections, and a symmetric network \cite{STA06, STA07}. Both are built by starting with four completely connected nodes of extremists
$F$ to which newcomers are progressively added; the original nodes influence the attitudes of others
but can also spontaneously return to the general population $G$. The level of extremism of each added
node depends on the attitudes held by its neighbors mimicking a $G \to S \to E \to F$ progression as originally 
proposed in \cite{CAS03}.  In the hierarchical network the attitudes of recently added nodes are influenced by those with more seniority, but the opposite is not true
resulting in a history dependent interactivity that represents groups with a rigid command-like structure. The symmetric network 
is more flexible so that all nodes mutually influence each other, regardless of when they joined.
The question is: to what degree will the original four fanatics influence incoming nodes?
What was found is that at steady state the hierarchical network is populated by
a uniform non-radical population made of $G$ individuals, although 
cohorts of low or moderate extremists, the $S$ and $E$ groups, can spread to large numbers before 
vanishing. In this case, the four original fanatics trigger a domino radicalization effect that can persist for a long time, but that 
eventually dissipates. This result is compatible with findings from the compartment model \cite{CAS03}
where a sufficiently large de-radicalization rate $\gamma_1$ yields a non-radical society at steady state. However, 
under the same initial conditions and for a limited $\gamma_1$ range, the symmetric network yields a much different scenario: at steady state  
comparable numbers of $G,S,E,F$ nodes arise, showing that the mutual reinforcement of attitudes, which is not possible in
the hierarchical network, greatly increases the spread and persistence of fanatic ideals.
Behaviors on the square lattice and comparisons with the compartment model are also discussed \cite{STA06}. 
These results confirm that connectivity structures and hierarchies
can greatly influence the spread of extreme attitudes, as already shown for
other epidemic phenomena evolving on networks of different types \cite{ALB02}.

\begin{figure}[t]
\hspace{-0.5cm}
\includegraphics[width=4.7in]{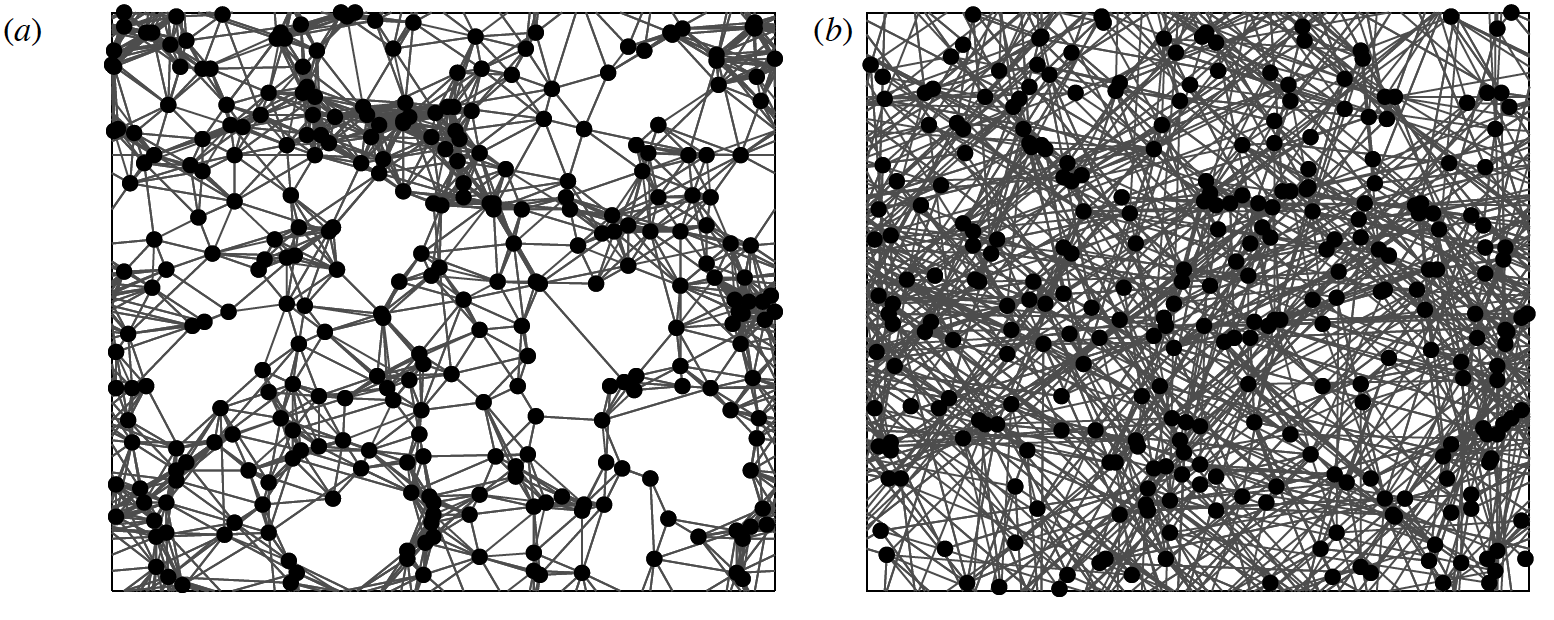}
\caption{Examples of networks where nodes are separated by a random distance $d$ and links
are probabilistically established drawing from a half-gaussian distribution 
$P(d;D) \sim e^{-d^2/ 2 D^2}$.  Here, $N=300$ and the average number of connections
per node is set at $\langle n \rangle =  8$. 
In panel (a) where $D=1$ the average distance between nodes $\langle d \rangle$ is 
less than in panel (b) where $D=10$.  The clustering coefficient $C$ is
defined as the average over all nodes of the number 
of locally connected triangles through a given node, 
divided by the maximal number of possible triangles the
given node can be part of. 
In panel (a) $C = 0.15$, in panel (b) $C=0.015$ revealing that clusters are more tightly connected
for lower values of $D$. These networks were originally used to study tuberculosis epidemics;
later, a $G \to S \to E \to F$ radical progression was implemented on them to study extremism. 
Taken from Ref.\,\cite{COH07}.}
\label{network}
\end{figure}

Networks used to study the evolution of diseases \cite{REA03, COH07} were later adapted
to describe the spread of radical religious ideologies \cite{CHE09}. Here, $N$
nodes are distributed in space at random so that each pair is separated by a socio-spatial distance $d$, which represents affinity,  
friendship or family closeness. The distance $d$ also determines whether nodes are linked or not:
connections are established probabilistically drawing from a half-gaussian distribution $P(d; D)$ of width $D$
so that the average number of connections per node is fixed at a given $\langle n \rangle$ and so that 
the average distance on the connected network is proportional to $D$, $\langle d \rangle \sim D$.
When three nodes make a three-way connection to form a triangle, they are
considered ``clustered." The cluster coefficient $C$ of a network is defined as the average number
of such triangles per node, over the maximal number of possible triangles that can be constructed
through it. As shown in Fig.\,\ref{network}\,(a),
small $D$ yields tight clusters of nodes with shorter links among them (large $C$, small $\langle d \rangle$) , 
large $D$ results in more global connections 
with relationships established across larger distances  
(small $C$, large $\langle d \rangle$) \cite{REA03}.
Once constructed, a radicalization process akin, but not identical, 
to the $G \to S \to E \to F$ sequence introduced in Eqs.\,\eqref{CC1}--\eqref{CC4} is imposed on the network.
One of the main differences with respect to the original compartment 
model in \cite{CAS03}, is that here fanatics are further divided in two classes: 
foot-soldiers $R_S$ and leaders $R_L$ which are associated to higher mortality rates
due to, for example, suicide bombings, counter-terrorism attacks or arrests. 
Each node $i$ is assigned an intrinsic ideology receptiveness $\tau_i \leq 1$, and a radicalization state $G,S,E,F$.
Progression along the radicalization hierarchy occurs with probability $1 - (1 - \tau_i)^k$,
where $k$ is the number of radical neighbors of node $i$, 
prioritizing those most likely to embrace the radical ideology.  Finally, the network is quasi-static 
as its structure changes only due to death whereby nodes are removed and 
reintroduced randomly, rearranging links. Simulations run on 
1,250 nodes, of which $10\,\%$ at $t=0$ were fanatics and the remainder part of the 
general, non-radical population, reveal that a sufficiently large initial number of fanatics can drive the core population
to radicalism before extinction.

Other modeling work considers networks where nodes carry a belief $B_i$ and 
are connected only if they are in close geographical proximity and their beliefs are similar.
These two elements model homophily, 
the well known sociological phenomenon of being drawn to those similar to oneself \cite{MCP01}.
Nodes adapt the average opinion of their neighbors
depending on a spread parameter $\alpha$, and on an intrinsic node vulnerability
$N_v$; the most influential neighbors are assumed to be the ones
with higher degree centrality \cite{NEW10}.  Finally, once a critical belief is achieved, the node is assumed to have radicalized
and will only interact with other radicals. Apart from scale-free networks, 
random and small-world networks are analyzed:  vulnerability and information spread are found to be more 
relevant than the underlying network structure in determining the 
abundance of radical nodes \cite{MEN18}. 

Some scale-free network models have studied radicalization 
as extreme forms of protest in the environmental and animal rights movements,
which at times may also lead to violent action \cite{PEN13}. The two movements
were analyzed together since they share many similarities in their
origin and ideals, yet display different characteristics in the way activists protest. Environmentalists gravitate towards demonstrative or confrontational actions, with
radicals opting for minor attacks on property; animal rights extremists display a greater 
tendency to target people \cite{SEE00, CAR12}.  
In the network model introduced in \cite{PEN13} 
nodes are assumed to carry a level of criminal propensity
and a susceptibility for changing their morality and self-control; these levels evolve depending
on social contacts with other nodes.  Empirical data collected by the authors
suggests that pre-radicalized animal rights extremists tend to create sub-groups who 
attract more of the same; peaceful protesters instead remain such even after joining an animal rights
movement. There is no equivalent correlation for environmental protesters.
Activist nodes in animal rights networks were thus assumed to preferentially link to nodes with the same criminality levels;
while environmental activists forged connections with nodes of the same morality level, mimicking homophily 
\cite{GAR10}.
Among the results is that initial conditions greatly affect the outcomes of activist campaigns, 
and that given the same initial conditions and network configuration, the final 
number of criminally-minded agents in animal rights 
movements is much higher than in the environmental case.
The social endorsement of opinions is also studied
in hierarchical Ising models, where opinions of individuals can evolve
continuously or via discrete jumps
\cite{MAA20}. The combination of opinion endorsement and discrete switching
allows for the coexistence of multiple stable states and the occasional emergence of 
polarizing opinions.

The structure of two historical, and purportedly scale-free networks, was also analyzed for 
the Cathar heresy in 13$^{th}$ century France -- the first organized challenge ever faced by the Roman Catholic Church -- 
and the Reformation in England in the mid 16$^{th}$ century that led to large scale conversions to Protestantism \cite{ORM04, ORM08}.
While the heresy in France was suppressed, the reformist movement successfully spread throughout England.
Although limited data was available, key players serving as network hubs were identified in both cases  \cite{ORM07}. By comparing the
two historical outcomes, the authors concluded that in order to stop extreme movements from permeating society,
it may be more effective to isolate key players on the network by neutralizing their close contacts, 
rather than risk creating cascade effects through active persecution, or even martyrdom. 
The authors argue that the lessons learnt may be applicable to current times. 

\begin{figure}[t]

\includegraphics[width=4.7in]{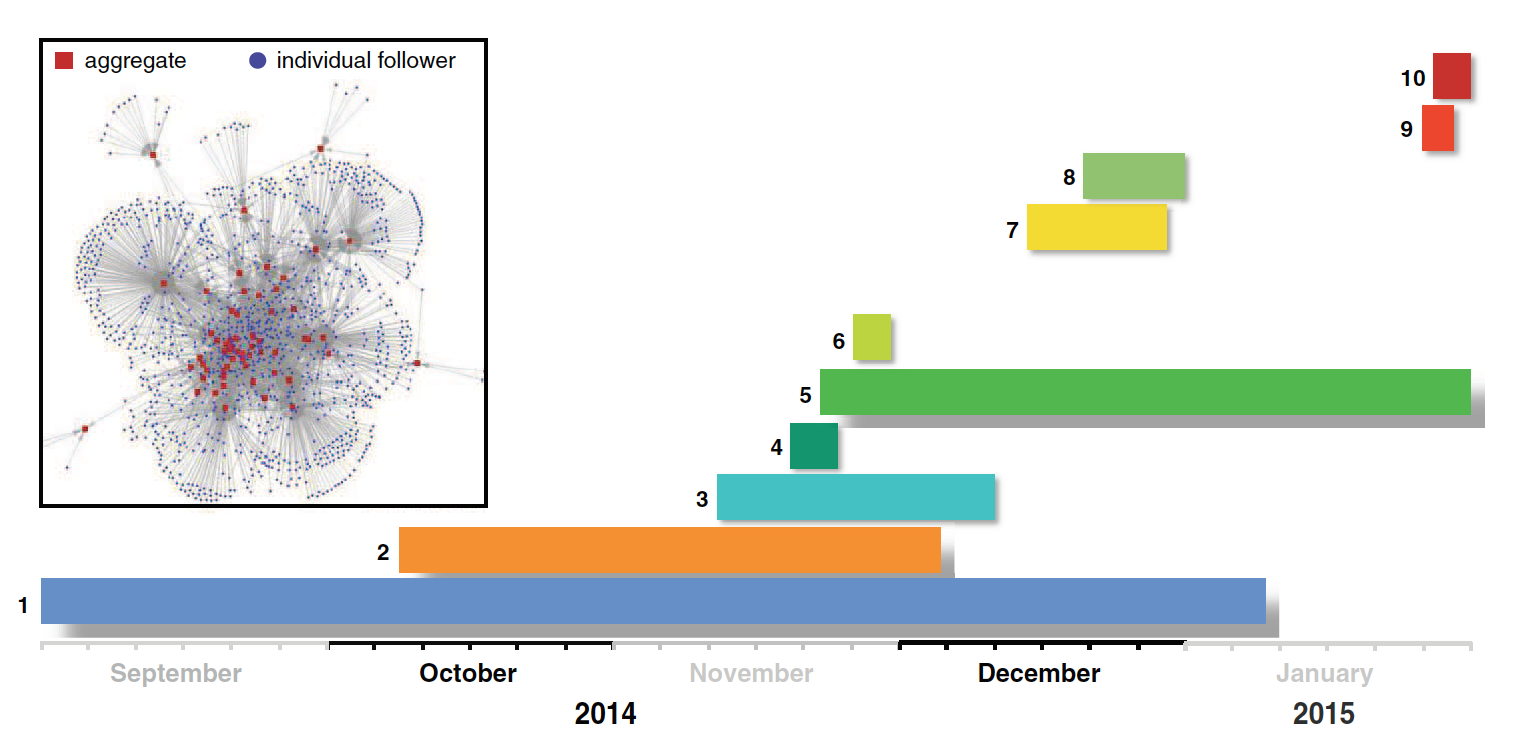}
\caption{Dynamics of ten pro-ISIS aggregate groups on the social network VKontakte. 
Horizontal bars shows the lifespan of each of these groups; the process of appearing and disappearing
is modeled as a coagulation-fragmentation process via Eqs.\,\eqref{coagfrag1} and \eqref{coagfrag2}. 
The inset shows a portion of an actual aggregate-follower network on January 1$^{st}$ 2015 
with individual followers (blue nodes) linking to pro-ISIS aggregates (red nodes). 
Taken from Ref.\,\cite{JOH16}.}
\label{coagfragfig}
\end{figure}

How followers are recruited or spontaneously join online extremist networks
was the subject of a detailed analysis performed on VKontakte \cite{JOH16}. 
This is a Russian-based online social networking service, with more than 350 million subscribers, 
heavily used by ISIS as a means to spread its
ideology in Chechen Russia. Roughly 196 pro-ISIS self-organized online aggregates 
(the equivalent of Facebook groups) were identified over an eight month period in 2015, 
attracting more than one hundred thousand adherents. The growth and decline of the size of these aggregates
were tracked over time revealing merging and shutdowns, possibly due to government 
or regulator intervention as shown in Fig.\,\ref{coagfragfig}.
The dynamics was then modeled as a coagulation and fragmentation process \cite{JOH16, RUS09} as follows

\begin{eqnarray}
\label{coagfrag1}
\frac{\partial n_s}{\partial t} &=& 
\frac{v_{\textrm{coal}}}{N^2}  \sum_{k=1}^{s-1} k (s-k) n_k n_{s-k}  - \frac{2 v_{\textrm {coal}}}{N^2} s n_s
\sum_{k=1}^{\infty} k  n_k - \frac{v_{\textrm {frag}}}{N} s n_{s}, \\
\frac{\partial n_1}{\partial t} &=& 
\label{coagfrag2}
 - \frac{2 v_{\textrm {coal}}}{N^2}  n_1 \sum_{k=1}^{\infty} k  n_k
 + \frac{v_{\textrm {frag}}}{N}  \sum_{k=2}^{\infty} k^2 n_k .
\end{eqnarray}

\noindent
Here $n_s(t)$ is the number of aggregates of size $s$ (i.e. comprised of $s$ individuals) at time $t$, 
and the total number of individuals is $N(t) = \sum_{s} s n_s(t)$. Individuals
are assumed to join and leave aggregates but not to exit the system, 
so that $N(t) = N$ is fixed and $n_s = 0$ for $s > N$. 
The three terms on the right hand side of Eq.\,\eqref{coagfrag1} refer respectively to: coagulation/merger 
between an aggregate of size $k$ and $s-k$ to form a larger
one of size $s$; the loss of an aggregate of size $s$ that merged with another one; the fragmentation or loss
of an aggregate due to its dissolution. Unaffiliated individuals belong to the $s=1$ aggregate. 
Rates of coalescence and fragmentation are given by $v_{\textrm {coal}}$ and $v_{\textrm {frag}}$ respectively. 
Eq.\,\eqref{coagfrag1} is complemented by Eq.\,\eqref{coagfrag2} which models
the dynamics for an aggregate of size $s=1$ and where no coalescence
is possible, since the minimum size of a coalescence process is $s=2$.
The two terms on the right hand side represent respectively the merging of an $s=1$ aggregate with any other,
and an increase in the number of $s=1$ aggregates due to the fragmentation of an existing
one. Using generating functions \cite{WIL90} it can be shown that at equilibrium 
the size distribution of aggregates $n_s^*$ scales according to

\begin{eqnarray}
\label{coagfragscale}
n_s^* &\sim&
\left( 
\frac{\nu_{\textrm {coal}}^{s-1} (\nu_{\textrm {coal}} + \nu_{\textrm {frag}})^s}
{ (2 \nu_{\textrm {coal} } + \nu_{\textrm {frag}})^{2s-1}}
 \right) s^{-5/2}.
\end{eqnarray}

\noindent 
In the limit of large $s$ a power-law distribution emerges $n_s^* \sim s^{- \gamma}$ where $\gamma = 2.5$, 
in good agreement with the $\gamma = 2.33$ seen in observations. Eqs.\,\eqref{coagfrag1} and \eqref{coagfrag2} 
also allow us to explore possible avenues of online anti-terrorism intervention. For example, 
focusing on the shutting down of smaller groups, before they become too potent is shown to be 
an effective strategy. Indeed, if the dissolution rate $v_{\textrm {frag}}$ is too slow, several clusters will rapidly 
coalesce into one super-aggregate. Finally,  the same authors used
a similar coagulation-fragmentation framework to describe the likelihood
that a terrorist attack will result in a given number of victims.  
Richardson's Law is a well known paradigm in political conflict, according to which 
the distribution of casualties due to violence follows a power law, 
so that the probability $p(x)$ of an event with $x$ deaths scales as $p(x) \sim  x^{-\gamma}$ \cite{RAP57}. 
The assumption made in \cite{BOH09, JOH15, CLA10}
is that an insurgent movement is made of various 
cells of extremists, each carrying an attack strength that represents the potential
number of casualties it may inflict, proportional to its size. 
Cells are continuously rearranging their attack strengths by coalescing and fragmenting
as described in Eqs.\,\eqref{coagfrag1}--\eqref{coagfrag2} so that the
attack strength distribution is the same as the size distribution
$n_s^* \sim s^{- \gamma}$ in Eq.\,\eqref{coagfragscale}. Thus, $n_s^*$ can be viewed as
a proxy for the number of casualties $p(x) \sim x^{\gamma}$. 
A convincing illustration of this parallel is that the scaling of the number of victims due to insurgent groups 
in Iraq, Colombia and Afghanistan
follows a power-law distribution, with exponents consistent with the
$\gamma \sim 2.5$ estimate predicted by Eqs.\,\eqref{coagfrag1}--\eqref{coagfrag2}.
Coagulation-fragmentation models can also be used to 
represent military and economic coalitions \cite{GAL23b}. 

\begin{figure}[t]
\includegraphics[width=4.0in]{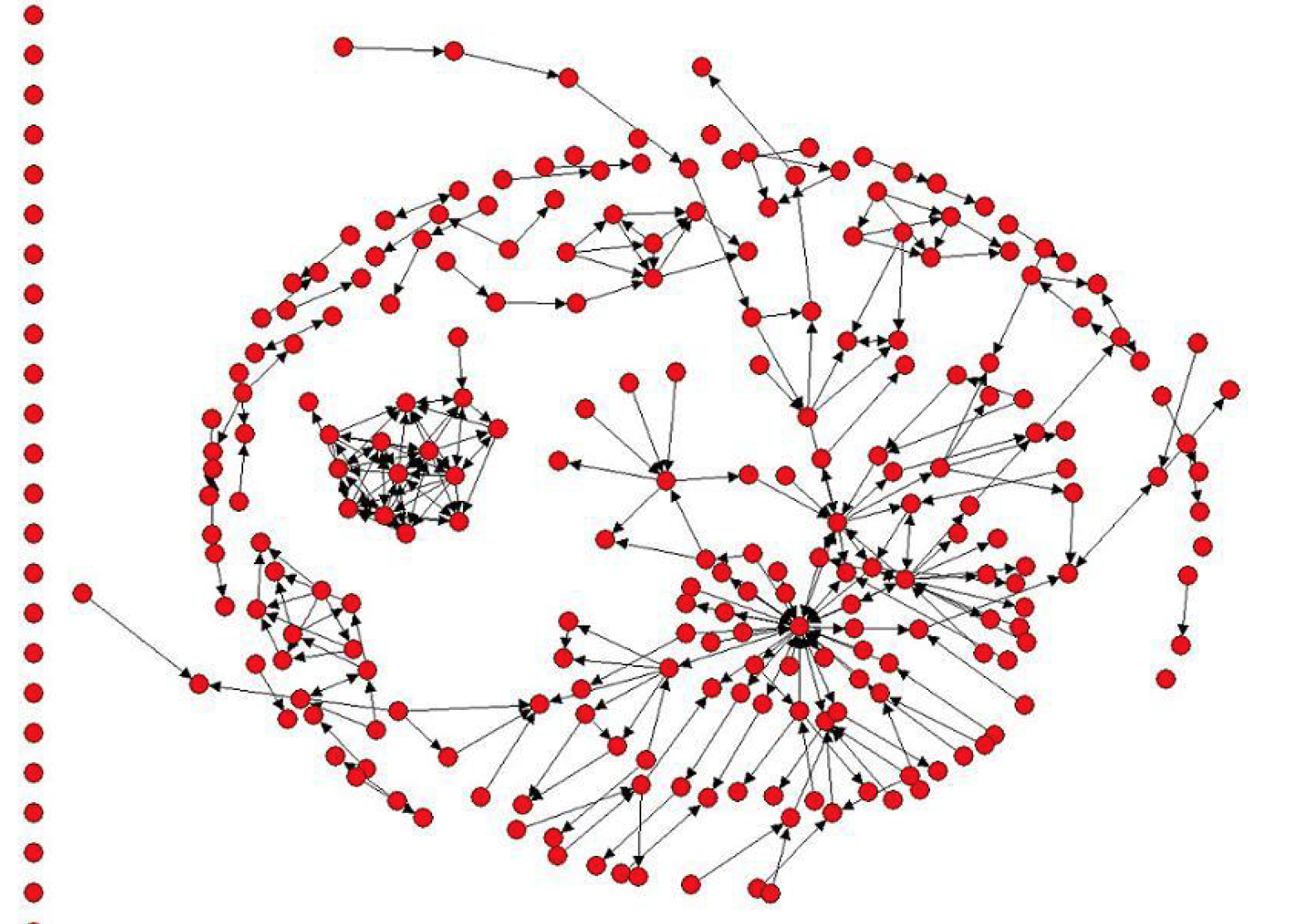}
\caption{An agent-by-agent Middle-East terrorist network visualized through the Meta-Matrix Text Analysis \cite{CAR07}. 
Media outlets searched to create this graph include major newspapers, magazines, journals, trial transcripts, book excerpts, 
scientific articles on Al-Qaeda in Iraq from 1977 to 2004. The nodes on the left are not directly linked to other actors. 
The isolated, circular sub-graph on the left side of the inner circle that is not connected to other agents  
represents individuals who were charged with the Khobar Tower Bombing in Saudi Arabia in 1994.
Relevant nodes were identified: one carries the highest cognitive demand (0.0915), degree centrality (0.0915) 
and betweenness centrality (0.0463); 
a distinct node 
has the highest clique count (12.0), and 
yet another distinct node has the highest number of simmelian ties (0.0282). Taken from Ref.\,\cite{DIE05}. 
}
\label{textnetwork}
\end{figure}

But how exactly is a network of operational terrorists structured? How does it change in time,
due to internal dynamics, changes to the socio-political environment, the need for 
secrecy, technological advances, and in response to counter-terror interventions \cite{SHA13}? 
Any progress in answering these questions may lead to a better assessment of terrorist hierarchies, and in 
developing policies and practices to best detect and disrupt them \cite{BER09}. 
Terrorist groups function underground and their secretive nature makes data collection challenging; to date, 
information has mostly been gathered via open-source texts from the media and publicly available legal transcripts
\cite{SAF17}.  
Initial efforts to understand terrorist organizations sprung in the early 1980s
to catalogue insurgent activities in Palestine, by the Provisional Irish Republican Army (PIRA), 
and covert operations by the KBG \cite{GIL14}. 
The September 11$^{th}$ 2001 terrorist attacks led to greater urgency in this direction; as much as possible, data collection, social network
and text analysis were used to identify emergent leaders, spheres of influence, 
hubs, graph structures of individual cells and of extremists frequenting web forums \cite{SAG04, KRE01, KOS06}.
Several computer-supported techniques combining network text analysis and
methods to classify organizational systems or social network analysis \cite{HEL11}
were introduced, such as the Meta-Matrix Text Analysis \cite{DIE04, DIE05},
the Dynamic Network Analysis \cite{CAR03, CAR06}, the 
Computational Analysis of Social and Organizational Systems \cite{CAR07}
the Counter-Terror Social Network Analysis and Intent Recognition \cite{WEI09}.
These packages evolved over time to allow for visual, sophisticated representations of 
connections of actors, locations, events, tasks, knowledge, resources, allowing
to study not just command hierarchies, but also relationships based
on economics or training.  Some software also allows for the removal of actors to forecast 
possible consequences to the terrorist structure \cite{CAR06} after their capture or death.
These features can be especially useful for diplomatic or military considerations.
One example of a network reconstructed 
via the Meta-Matrix Text Analysis is shown in Fig.\,\ref{textnetwork}.

What emerges is that terrorist networks tend to display a highly centralized organization, but counter-terror pressures 
compel them to modify their structures and actions, forcing them to find compromise between
security and communication \cite{RES06, KNO15}. Typically, leadership roles are re-adjusted and
hierarchical organizations become more decentralized so that the network may survive a negative environment.
This trend is observed in many networks, including in al-Muhajiroun, a British group that 
advocates the establishment of an Islamic state in the country \cite{KEN13}:
once its leader left Britain, followers adapted their operations, 
created spin-off groups, kept recruiting new members, and shifted towards online communication.
A similar trend was seen in the militant Islamic group led by Noordin Top in Indonesia between
2000-2010: as authorities tried to disrupt the organization, its topography changed
in terms of centralization and density \cite{EVE13, EVE15}.  Similarly,  Al-Qaeda abandoned its initial corporate-like command structure 
in favor of a so-called leaderless jihad \cite{SAG08} due to increasing scrutiny after the 
September 11$^{th}$ terrorist attacks. 
These organizational changes are also confirmed by the text analysis methods described above \cite{CAR06},
where over the course of a decade, the density and communication levels within Al-Qaeda cells were seen to decrease.  
An evolutionary process to describe Al-Qaeda's path from a tightly coupled organization, to a coupled network,
to loosely coupled movement is illustrated in \cite{JAC06}. Furthermore, in its post September 11$^{th}$ incarnation, 
Al-Qaeda has been described as a ``dune-like" organization \cite{MIS05, KIR07, JOR08} whereby its affiliates are encouraged to operate 
independently and form ``self-starter" cells while still part of its broader network. Among the many drawbacks to this tactic
is that rebel groups may splinter and antagonize the once centralized command.
This is the genesis of present-day ISIS that began as part of the umbrella
of organizations under Al-Qaeda's influence, but that rapidly emerged as one of its strongest contenders.

Since the few, but well connected, nodes typical of scale-free networks may jeopardize the operational
security of terrorist cells, it is important to understand which organizational structures best serve
covert organizations. The tradeoff between operational efficiency and secrecy in terrorist networks was analyzed 
in \cite{LIN09} where various models of communication on connected graphs were compared and contrasted
under different detection risk scenarios. In this work, each node $i$ of a $g$ graph
is characterized by a degree $d_i$. The total number of nodes is $n$ and the total number of links
is $m$. First, a communication measure $I(g)$ is defined based on the connectivity of the graph \cite{NEW10}

\begin{eqnarray}
\label{Ig1}
I(g) = \frac {n(n-1)}{\sum_{ij} \ell_{ij} (g)},
\end{eqnarray}

\noindent
where $\ell_{ij}$ is the shortest distance between nodes $i$ and $j$, and $n (n-1)/2$
is the number of pairs in a network of $n$ nodes. The quantity $I(g)$ can be thought of as the normalized reciprocal of the total distance
and as a result $0 \leq I(g) \leq 1$.
Since $\ell_{ij} \to \infty$ for unconnected nodes, $I(g) \to 0$ for a disjoined graph;
$I(g) = 1$ on a fully connected graph.  Two factors are assumed to contribute to the security of node $i$: 
the exposure probability of being identified as a member of the network, $\alpha_i$,
and the fraction of the network that remains unexposed after a given node has been detected,
$u_i$.  The overall likelihood of the network remaining unexposed is thus given by 

\begin{eqnarray}
\label{Sg}
S(g) = \sum_{i} \alpha_i u_i.
\end{eqnarray}

\begin{figure}[t]
\includegraphics[width=4.0in]{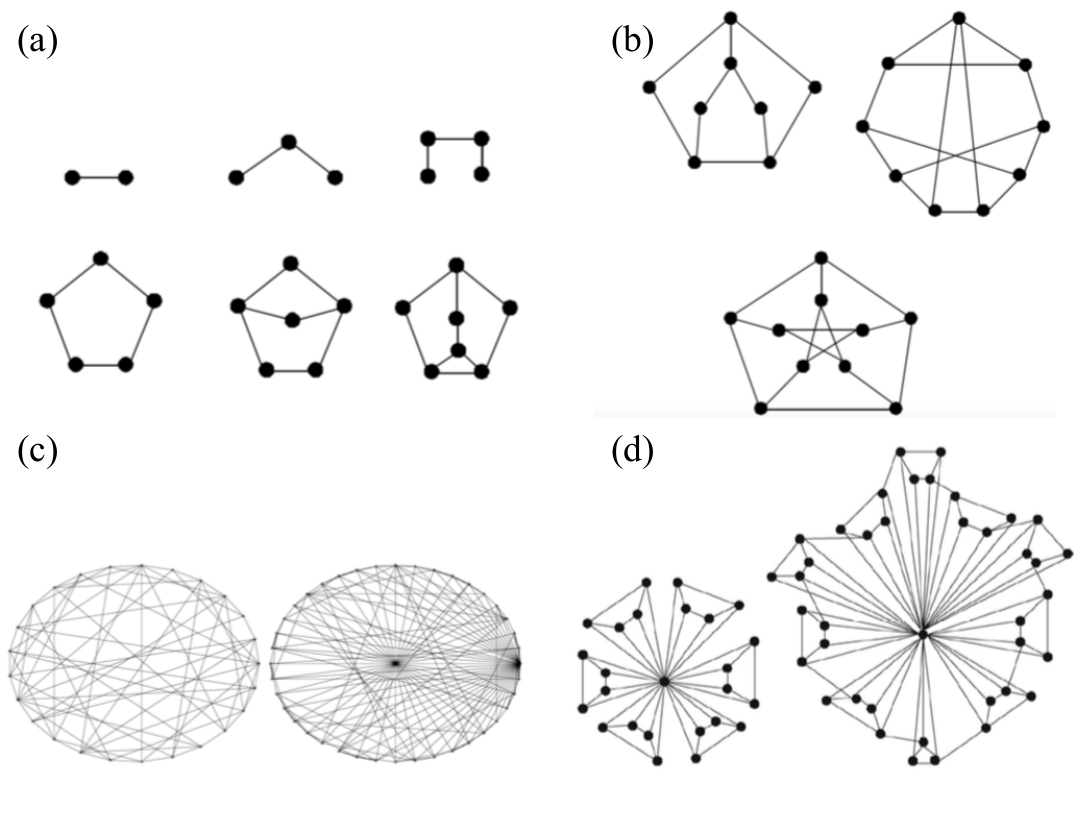}
\caption{Structures of covert graphs leading to the maximal information and secrecy measure
$\mu(g) = I(g) S(g)$ defined via Eqs.\,\eqref{Ig1}--\eqref{Sg}. In Eq.\,\eqref{Sg} node detection is
proportional to its centrality in the network, $\alpha_i = (d_i+1)/ (2 m+n)$, 
and full exposure upon detection $u_i = (n - d_i + 1)/n$ is assumed.  
Panel (a) shows the optimal graphs for $n=2$ through $n=7$, and panel (b) displays  
approximate optimal graphs obtained via simulations for $n=8,9,10$.
Different optimal structures arising from different information measures are shown in 
the lower panels.  Using the same $\alpha_i, u_i$ as in panels (a) and (b), 
panel (c) shows approximate optimal graphs 
for $n=25$ and $n=40$ under $I(g)$ given by Eq.\,\eqref{Ig1}, 
whereas in panel (d) $I_2(g) = 1/ D(g)$ is used for the same number of nodes, 
$n=25$ and $n=40$ and where $D(g) = \max_{\{i,j\} \in g} \ell_{ij}$ is the diameter of the network. 
Taken from Ref.\,\cite{LIN09}.}
\label{Nashpics}
\end{figure}

\noindent  
A terrorist organization typically aims to enhance its
information communication $I(g)$, while simultaneously maintaining high secrecy levels
$S(g)$ for security purposes. This trade-off motivates the search for an optimal graph 
that maximizes the product of these two indicators $\mu(g) = S(g) I(g)$. The first scenario analyzed 
is that of a uniform exposure probability $\alpha_i = 1/n$ and 
where, if node $i$ is uncovered as part of the network, all others connected to it become exposed
as well with a link detection probability $p$, so that $u_i = (n - p d_i - 1)/n$. Here, $p=1$ represents the case where all links 
connecting node $i$ to others are exposed; $p=0$ indicates an impenetrable network where the identification of one
node still allows for the other $n-1$ to remain covert. 
The other scenario is that of node detection being proportional to its centrality in the network,
$\alpha_i = (d_i+1)/ (2 m+n) $ and where a detected node leads to all its connections being exposed
$u_i = (n - d_i - 1)/n$.  Game theoretic Nash bargaining is then used to determine
which $g$ structure leads to maximal $\mu(g)$ for small values of $n$ \cite{NAS50, MAR98}; simulations
are used in the case of larger $n$, where $\mu(g)$ is explicitly evaluated over multiple
configurations and the maximum value recorded. 
In the first case, if the link detection probability is high, $p \to 1$, 
the Nash bargaining solution is a star graph. As the link detection probability $p$ is decreased,
a transition is observed and a complete graph becomes best. 
In the second case, cellular networks such the ones shown 
in Figs.\,\ref{Nashpics}(a)--(c) are optimal.  As a result, optimal covert networks should incorporate all-to-all communication when 
detection risks are low; a star configuration when communications can be intercepted but nodes are equivalent; 
reinforced rings, wheels and other 
cellular structures if centrality is important.  Other choices of $I(g)$
may lead to different optimal graphs; for example in 
Fig.\,\ref{Nashpics}(d) best structures are shown if the communication measure is set at $I_2(g) = 1 / D(g)$ 
where $D(g)$ is the diameter of the graph defined as the 
maximal distance among all nodes, $D(g) = \max_{\{i,j\} \in g} \ell_{ij}$.
Among the many variants of this work, one of the most notable is 
the inclusion of node heterogeneity where a specific interaction between 
a pair of nodes may be riskier than others. In this case network structure is optimized by 
having the pair in question being the least connected to the remainder of the network. 
This theoretical framework was later
applied to the 2002 Bali attacks by Jemaah Islamiya, an Indonesian Al-Qaeda affiliated Islamist group,
for which prior knowledge of operational and command nodes was known \cite{KOS06}.
Links were assigned heterogeneous weights representing different risk levels. The 
optimal networks for $n=25$ and $n=40$ consist of cellular structures around a centralized individual, less dense than those
observed in Fig.\,\ref{Nashpics}(c) since the high risk link carries limited connectivity to the remainder
of the network \cite{LIN09,HUS13}. 

The structure of actual operative cells that attacked 
five South East Asia locations and Madrid, Spain was analyzed in Ref.\,\cite{HEL11}.
It was found that all 
terrorist networks became increasingly dense and cohesive as 
they prepared for executing the above attacks; these networks did not display scale-free characteristics,
nor did they seem to reorganize much to optimize communication secrecy.
This has led to the interesting theory that as they prepare for action, 
and especially if they operate under a sense of security, terrorist groups
will not devote resources to optimize their internal structure, as it 
would distract from their main objective, which is to attack. It is only the threat of detection and the need 
for covertness that catalyzes changes to existing patterns,
since links that were forged over time, not necessarily in an organic way and connecting individuals with different
ideals, skills and personalities, require costly, time-consuming intervention to rearrange. 
As a corollary, it is suggested that a possible counter-terrorism strategy could be
to foster a false sense of security, allowing for connections to densify over time 
so that once intervention occurs the maximal number of members of the network will be exposed.
The same authors propose that rather than eliminating hubs of well-connected individuals
hoping that the network will collapse, a better strategy in the absence of a scale-free
structure, or in the presence of networks that regenerate themselves, is to 
employ enough resources to target the entire terrorist cell.  The case of terrorists organized in a
clique of $N$ nodes, a graph where every two nodes are directly connected, has been analyzed
in \cite{LIN08}. Because of the structure of a clique, the capture of any of its members would expose all. 
If detection probability is set at $p$ per link, results from percolation theory imply that 
cliques of size $N \geq 1/2 p^2$ will be detected with high probability. 
This non-linear outcome suggests that any investment to
increase surveillance to $p$ would translate to capture rewards of order $p^2$; seen from 
a different perspective this result implies that to avoid detections, as $p$ is increased, 
cliques must shrink by $p^2$ members.  Other work addresses the question of
how hierarchical terrorist groups would function when crucial members
are captured or killed \cite{FAR03} and advocates for searching network cut-sets,
that is nodes whose removal disconnects leaders from followers.  

Dynamical shifts between 1989 and 2003 in the structure of the Global Salafi Jihad (GSJ) terrorist group, a Muslim revivalist
movement, was studied using open source information \cite{JIE09}. Results
are consistent with the dynamical changes described 
above: the GSJ network was observed to be rather random at the onset, 
but acquired a scale-free topology as new members joined while others left, due to pressure
from counter-terror operations.  Average degree, degree distribution, and network diameter, defined as the average length of
the shortest path between any pair of nodes, were used as statistical measures. 
Vulnerability to random failures, targeted attacks, and counter-terror
efforts were all analyzed by simulating node removal; the loose network was found to be
resilient to all disruption efforts.  

The Stochastic Opponent Modeling Agent (SOMA) was originally developed to understand 
how social, economic or religious communities behave; later it was specifically applied to 
terrorist organizations \cite{SUB07a, SUB07b, SLI08}. Datasets tracking rebellion 
and protest movements worldwide are mined to generate rules of behavior for known 
violent ethno-political or ethno-nationalist groups \cite{ASA07, MAR05}.  These rules are associated
to given actions such as kidnapping or orchestrating transnational attacks, and to various
environmental characteristics, such as receiving financial or military 
support from a foreign country. From the data, one can then reconstruct likelihoods for a given organization to 
execute a given action under given conditions.  This is valuable since the number of action-condition combinations 
is very large and a human analyst could easily miss interesting hypotheses.  
As an example, Hezbollah,  which is designated as a terrorist group by the United States, 
was tracked for 23 years and it was found that when involved in inter-organizational conflicts, it is less likely to engage in transnational violence  \cite{MAN11a}.
The SOMA framework was also used to study terrorist attacks on Israeli citizens by Hamas between 1987 and 2004 \cite{MAN08}. 
Results show that amidst a background of consistent violence against Israel, 
Hamas' capabilities or the political landscape may help explain why certain periods
were marked by especially high levels of violence.  
For example, Hamas appeared to especially engage in violent activities in years where it also provided social
services to Palestinians, arguably having used these services to build a stronger recruitment base. 
Similarly, its kidnapping of Israeli citizens increased at times of heightened 
conflict with Fatah, its major rival in the area, purportedly to raise its status in the area.
The study predicted that as Hamas expanded its power in Gaza, the likelihood of attacks against Israel would increase.
 Other contexts in which SOMA-rules were applied include understanding the behavior of actors in the
Afghan drug economy \cite{SLI07} and of separatist  movements in the Jammu and Kashmir provinces of India and Pakistan \cite{MAN11b}.
These findings do not explain why certain correlations are observed, however they
are still useful as they pose new questions and offer new avenues of investigation.

The topological role of women in terrorist organizations is explored in  \cite{MAN16};
data mining performed on the social media outlet VKontakte and on offline
records detailing the 1970-1988 activity of the Provisional Irish Republican Army (PIRA)
show that in both organizations women occupy nodes with superior network connectivity, guaranteeing robustness and 
survival to the network itself \cite{MAN16}. Betweenness centrality is a good indicator to measure how influential a given node is
in a network \cite{NEW10}; it is defined as the fraction of shortest paths connecting any two nodes that pass through a given node. In a network
of covert extremists, understanding shortest path features is especially relevant,
since adding any extra step to connect players represents risk and potential cost. The analysis in \cite{MAN16} 
shows that betweenness centrality on female nodes is much higher than on male nodes. 
The high interconnectivity of women implies that they 
play important roles as communicators of messages and goods; it would be interesting to include this feature into 
future dynamical models of terrorist networks. 

Finally, in order to succeed when planning and executing attacks, terrorist groups
may require the support of passive sympathizers who ensure secrecy and offer logistical support. 
This aspect of terrorism has been studied through percolation models on lattices \cite{GAL02,GAL03a,GAL03b}
whereby if the number of sympathizers exceeds a percolation threshold, terrorist operations are considered likely to succeed. 
A social dimension is included via socioeconomic, political, or religious discontent through which
individuals establish connections. Within this framework, given a fixed number of sympathizers, 
thwarting attacks depends on raising the percolation threshold; 
this can be achieved by decreasing connectivity between individuals, or by reducing the 
social dimension that shapes these connections. Governments that do not wish to imprison passive sympathizers, or
restrict the social connectivity of its citizens, may instead opt to disentangle intricate societal issues so that
they do not unite large numbers of dormant, unconnected sympathizers, which at the most
will form local, unconnected clusters.

\section{Game theoretic models of terrorism}
\label{gametheory}

In previous sections we discussed the radicalization of individuals,
and how new recruits join terrorist organizations. We also reviewed
the social structures of terrorist networks that facilitate communication while evading detection, and how various cells
may merge or dissipate.  All these human and material resources
determine the organizational capability of a terrorist organization; 
whether such capabilities translate into a successful execution
however, depends on how extremists interface with 
existing deterrents. Before striking, terrorist organizations must rationally weigh their probability for success, and evaluate their
expected gains against potential costs and the risk of being exposed. 
Attacks may thus be executed in the immediate, postponed to future times, 
or aborted.  The process of rational decision-making 
is best explored through the tools of game theory \cite{OSB94, TAD13}.
Aerial hijacking and hostage taking incidents, 
which were common in the 1980s 
were among the first terror events to be described via game theoretic models 
\cite{COR81,ATK87,ZEE87,LAP88,WIL00}. 
Despite the fanatical ideologies involved,  hijacker reactions and 
negotiating strategies with counter-terror agents
were found to be remarkably rational. 
This was interpreted as 
a consequence of both sides seeking to predict future developments
in a volatile situation, so as to avoid unexpected occurrences \cite{RUB86}.
Thus, differential games appear to be particularly apt to model the dynamics between
terrorists and counter-terror agencies, to help estimate the risk
of attacks, and to assess prevention protocols
\cite{MAJ02, KAP05,BEH07,CAU08,FEI08,CAU09,NOV10,FOK13,MEG19a,MEG18b}. 

Differential terrorism games originate from dynamical compartment models \cite{UDW06, KAM06}. 
Here, extremists and governmental agencies adjust the intensity
of their activities as they interact with each other and other sociopolitical entities, 
seeking the most favorable outcome. The strength of a 
terrorist organization is described by a scalar state variable $x (t) \ge 0$, 
which represents one or a combination of resources the organization possesses, 
such as the number of recruits, weapons, supporters from the general public, 
financial capital, knowledge and access to new technologies \cite{KAP05,BEH07,KEO03}. 
The state variable $x$ evolves over time as follows
\cite{FEI08,NOV10,MEG19a,MEG18b}

\begin{equation}
   \frac{\mathrm{d} x}{\mathrm{d} t} = I (x; \mathbf{u}) - O (x; \mathbf{u}) \equiv {\mathcal C}  (x; \mathbf{u}),
   \label{EQ:GAME}
\end{equation}

\noindent
where $I$ and $O$ represent the influx (accumulation) and 
the outflux (consumption) of resources, respectively, and $\mathbf{u} (t)$ denotes an
array of time-dependent control parameters that reflect strategy adjustments
of both terrorists and governmental agencies.
Typically, the influx term $I$ is assumed to be independent of
external circumstances and to grow linearly at rate $r$,  $I (x)= r x $ \cite{NOV10,MEG19a,MEG18b}. 
Some models include a carrying capacity on the amount of resources an organization can maintain,
and adopt a logistic growth $I (x) = r x (x_\mathrm{max} -x)$, where $x_\mathrm{max}$ specifies the
maximum possible resource level \cite{BEH07}. A constant influx is sometimes included to ``jump start''
the model and usually set to a very small value \cite{CAU08,CAU09}. 
Further dependence on the control parameters $\mathbf{u}$ may be included.
For example, an overly aggressive counter-terrorism policy may antagonize the general public, 
boosting public support for terrorist groups, and favor the recruitment of new members \cite{KAP05,BEH07,CAU08}.
The most simplistic assumption is for the outflux term $O$ to depend  
on a two-component control parameter $\mathbf{u} (t) = (u_1 (t), u_2 (t))$ 
where $u_1 (t) \ge 0$, $u_2 (t) \ge 0$ represent the intensities of terrorist
activities and anti-terrorism operations, respectively, so that \cite{NOV10,MEG19a,MEG18b}

\begin{equation}
   O (x; \mathbf{u}) = h(u_1, u_2).
    \label{EQ:OUTFLUX}
\end{equation}

\noindent
The function $h$ is referred to as a ``harvest''; it specifies the consumption and loss 
of terrorist resources due to attack executions,
arrests, foiled attacks, and other events. 
It is generally assumed that $\partial h / \partial u_i > 0$, for 
$i =1,2$ so that increases in terrorist and governmental activity 
result in increases to resource harvesting. Further assumptions
can be made on the second-order partial derivatives of $h$. Imposing 
$\partial^2 h / \partial u_1^2 > 0$
indicates that as the intensity of terrorist activities increases, their consumption of 
resources intensifies; $\partial^2 h / \partial u_2^2 < 0$ 
indicates that as the intensity of counter terrorism activities increases,
terrorist resource consumption decelerates. Here, the
underlying idea is that as counter-terrorism measures
intensify, decreased gains are to be expected in terms of
losses to terrorists. Finally, 
$\partial^2 h / (\partial u_1 \partial u_2) \ge 0$ indicates that 
increased attack intensities lead to greater anti-terrorism operations,
whereas increased anti-terrorism activities will increase terrorist consumption of
resources. Lastly, $h$ satisfies the boundary conditions
$\lim_{u_i \rightarrow 0} h(u_i, u_j) \rightarrow 0$, and
$\lim_{u_i \rightarrow \infty} h(u_i, u_j) \rightarrow \infty$
for $i, j = 1, 2$. These are known as the Inada conditions \cite{INA63}: there is no 
spending of resources if there are no terrorist attacks nor any anti-terrorism measures in place; 
when either terrorists or governmental agencies 
strongly increase their activities,  the harvest of resources becomes infinitely large.
Some models include natural attrition to the outflux of resources,
such as departure of personnel and expiration of supplies, which limits
resource growth \cite{CAU09}.
Both terrorists and anti-terrorism authorities aim to 
maximize given utility functions by varying their respective control parameters $u_i$
for $i=1,2$ under the constraints given by Eq.\,\eqref{EQ:GAME}.
The terrorist utility function is written as

\begin{equation}
   J_1 (T) = \int_0^T e^{- \rho_1 t}         \left[ \alpha x (t) + \beta u_1 (t) \right] \mathrm{d} t
         + e^{- \rho_1 T} \left[ \alpha x (T) + \beta u_1 (T) \right].
        \label{EQ:J1}
\end{equation}

\noindent
Eq.\,\eqref{EQ:J1} implies that terrorists benefit from increasing 
resources $x$ and attack intensities $u_1$; 
$\alpha > 0$ and $\beta > 0$ define their relative weights, respectively.
The exponential pre-factor indicates that resources and attack capabilities decay 
in time at rate $\rho_1$.  It is important to note that 
$\rho_1 \leq r$ leads to unrealistic scenarios, whereby the best strategy for the
terrorist group is to accumulate resources and capabilities
\textit{ad infinitum} without ever attacking. Hence $\rho_1 >  r$ is always implicitly assumed.
The termination time $T > 0$ specifies the
time horizon over which the optimization game is to be conducted. 
For example, $T$ may be a 
presidential term if the terrorist organization aims to undermine the
incumbent administration; incumbents may instead seek political capital 
through counter-terrorism measures as they strategize for the next electoral cycle
\cite{BEH07}. If no specific timeframe is considered, $T \to \infty$, so that the
second term in Eq.\,\eqref{EQ:J1} vanishes. 
Governmental agencies seek to maximize 
the loss of terrorist resources, the $h(u_1, u_2)$ harvest, and to 
minimize $x$ and $u_1$ using as little effort $u_2 $ as possible. 
These assumptions lead to the following government utility function

\begin{eqnarray}
   J_2 (T) & = & \int_0^T e^{- \rho_2 t}
         \left[ \gamma h(u_1(t), u_2(t)) - \kappa x (t) - \sigma u_1 (t) - \eta u_2 (t) \right] \mathrm{d} t
         \nonumber \\ 
         & & + e^{ - \rho_2 T}
         \left[ \gamma h(u_1(T), u_2(T)) - \kappa x (T) - \sigma u_1 (T) - \eta u_2 (T) \right],
         \label{EQ:J2}
\end{eqnarray}

\noindent
where $\gamma > 0$, $\kappa > 0$, $\sigma > 0$, and $\eta > 0$ are constant weights.
Just as for $J_1(T)$, in order for realistic scenarios to emerge $\rho_2 > r$: if this condition is not obeyed, 
the best strategy for governmental agencies would be to allow extremists to
keep accumulating resources and attack capability,
which is undesirable. 

The simplest setting for this two party game is to assume that terrorist organizations and 
authorities do not wait for each other's actions to establish
their own tactics. The model is thus an open-loop game. 
To proceed, for simplicity, one can set $T \to \infty$ so that $ J_1 = \int_0^{\infty} j_1(t) dt $ where
$j_1(t) =e^{-\rho_1 t } (\alpha x (t) + \beta u_1(t))$; 
similarly for $J_2$. To calculate the optimal ${\bf u }(t)$ under the constraint given by Eq.\,\eqref{EQ:GAME}
a Lagrangian function ${\mathcal L}_1(x, \lambda_1, {\bf u}, t)$ is derived 
by extending the $J_1$ payoff to incorporate the constraint. 
This is done through a continuum of Lagrange multipliers $\lambda_1 (t)$ resulting in

\begin{eqnarray}
{\mathcal L}_1 =  \int_{0}^{T} j_1(t) + \lambda_1 (t) ({\mathcal C}(x;{\bf u}) - {\dot x}(t)) dt.
\end{eqnarray}

\noindent
The quantity $H_1(x,\lambda, {\bf u}, t)  = j_1 (t) + \lambda_1 (t) {\mathcal C} (x; {\bf u})$ 
is referred to as the Hamiltonian; the corresponding $H_2(x, \lambda_2, {\bf u}, t)$ can be derived for the
utility $J_2$, subject to the same constraint. 
Using Pontryagin's maximum principle the optimal control ${\bf u} =(u_1, u_1)$ and the optimal trajectory $x(t)$
can be found by minimizing the $H_1, H_2$ Hamiltonians \cite{PON62}. 
The resulting Nash equilibria depend on the form of the harvest function $h(u_1, u_2)$:
if $h$ increases at a relatively steep rate as a function of $u_1$, the optimal strategy is
for terrorists to attack less and $u_1$ is kept low; if counter-terrorism operations 
strongly affect terrorist activity, the best strategy for governmental agencies
is to increase $u_2$.

In more realistic settings, one side observes the actions of the other and
adjusts its strategy accordingly,
For example, the terrorist organization may secretly observe how
the government deploys its resources 
before choosing whether and which target to attack; similarly once counter-terrorism
measures are in place, authorities may wait for terrorist organizations to strike first, 
and later adjust their responses. When one player is a leader and the other is a follower in a two-player game,
the model is a closed-loop Stackelberg competition, for which sub-game perfect Nash
equilibria may exist \cite{WAT13}. Such equilibria can be found by solving the optimal strategy
of the follower $u_\mathrm{F}^*$ as a function of the leader's $u_\mathrm{L}$;
the leader determines its optimal strategy $u_\mathrm{L}^*$ accordingly.
Note that the Hamiltonians are different in the open-loop and in the Stackelberg games.

Regardless of which player takes on the role of the leader,  
the Stackelberg game always results in higher payoffs to the leader
than in the open-loop game. This signifies that, given the same circumstances
and model parameters, a good strategy for either terrorists or anti-terrorism agencies 
is to be proactive and act first. When terrorists are the leaders, 
both groups act more cautiously than in the open-loop scenario, there are 
fewer terrorist attacks and fewer counter-terrorism operations, 
but more resourceful terrorists at the end of the game \cite{NOV10}. 
If governmental agencies are the leaders, optimal strategies on both sides depend
 on $\sigma$, the weight assigned to $u_1$ in Eq.\,\eqref{EQ:J2}. 
This quantity can be interpreted as the damage inflicted on governmental assets per terrorist attack. 
For relative small $\sigma$, $x (t)$ becomes negative at long times, and a Nash equilibrium does not exist. 
For moderate $\sigma$, when damages inflicted are modest, the equilibrium $u_2$ value is smaller,
and $u_1$ larger than in the open-loop game: counter-terrorism agencies are more guarded, 
and this allows terrorists to accumulate more resources.  
Finally, for large $\sigma$, when attacks are on a larger scale, the equilibrium $u_2$ is larger
and the equilibrium $u_1$ is smaller in the Stackelberg than in the open-loop game:  
counter-terrorism agencies are more proactive in preventing 
these large scale attacks, and extremists become more cautious. 
Eq.\,\eqref{EQ:GAME} may be extended to explicitly include public support of the government, which 
may grow in time, decrease due to terrorist attacks, or be boosted by successful  
anti-terrorism operations \cite{MEG19a}.
The two players may also play a zero-sum
game, where one seeks to minimize the payoff of the other,
rather than maximize its own utility \cite{MEG18b}. 

Excessively aggressive anti-terrorism tactics 
that can alienate public opinion and increase 
support for terrorist organizations have also been investigated 
\cite{BEH07}. Here, $I(x; {\bf u})$ is augmented by 
$I_\mathrm{ant} (x; {\bf u}) \propto u_2(t)^2$, representing
unintended advantages that government efforts yield to terrorists, 
limiting their own intervention capability. In most scenarios, 
a perpetual tug emerges between the two parties: by provoking 
authorities to respond overaggressively, terrorist attacks may self-sustain themselves
despite the large consumption of resources, and a cycle of violence emerges.
This scenario is avoided if the proper values of 
$\alpha, \beta$ are selected in Eq.\,\eqref{EQ:J1},
in particular if the terrorist group is more interested
in the collection of resources (large $\alpha$), and less in the
execution of attacks (low $\beta$).  It is thus suggested that 
governmental agencies should attempt to influence terrorists to 
pursue political goals rather than conduct destructive acts; this shift
may also emerge naturally as extremist groups mature. 
Historically, many former terrorist organizations did transition into legitimate political parties,
and at least to some degree renounced violent methods. Well known examples include the Revolutionary 
Armed Forces of Colombia (FARC), the Provisional Irish Republican Army (PIRA), and at least
partially,  the Palestine Liberation Organization (PLO).
This conclusion is similar to what observed in Ref.\,\cite{SHO17} and discussed
in Sec.\,\ref{compartment},  where using a different game-theoretic mathematical framework, 
it was shown that the optimal strategy for a small radical faction to grow after having established itself through
violence, is to gradually disengage from it and focus on indoctrination. 

The effects of government authorities applying negative (``sticks'') and positive (``carrots'') 
counter-terrorism incentives is analyzed in \cite{BIE11} under the assumption that negative incentives may be
detrimental causing, for example, the emergence of hatred in the local population. 
A related ``fire or water'' model was also developed whereby two 
different anti-terrorism strategies can affect
the strength of a terrorist group $0 \leq x \leq 1$  \cite{CAU08}. Fire strategies to contrast $x$
are cheaper, more aggressive, and more effective but less precise; 
they are also more controversial  and may indirectly advantage terrorist groups.
Water strategies are less invasive, precise and surgically planned,
and remove terrorist threats without antagonizing the general public. 
They are also assumed to be less effective and more costly than
fire strategies.  The objective of anti-terrorist authorities is to minimize terrorist strength $x$,
as well as expenditures for water $u$ and fire $v$ strategies, subject to the
constraints $u, v \ge 0$; reactions from terrorists and related costs are disregarded.
As a result, $\sigma = \gamma = 0$ in Eq.\,\eqref{EQ:J2}, and
terrorist responses $u_1$ and the harvest $h$ are not included in the expression
for the $J_2$ government utility function. 

\begin{figure}[t]

\includegraphics[width=4.0in]{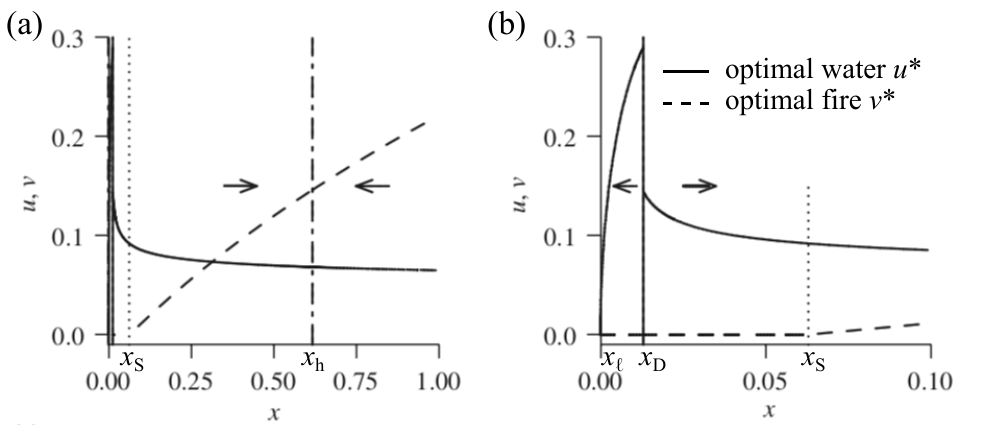}
\caption{Optimal control values of water $u^*$ (solid line)
and fire $v^*$ (dashed line) counter-terrorism strategies
as functions of terrorist strength $0 \le x \le 1$. 
Two non-zero equilibria $x_{\rm h}$ and $x_\ell$ arise. 
Panel (a) shows
the emergence of the high-level Nash equilibrium value $x_{\rm h}$,
with the horizontal arrows depicting the phase flow.
For $x > x_h$ fire strategies weaken terrorist activity whereas
for $x < x_h$ they embolden them.
Panel (b) shows the magnified range
$0 \le x \le 0.1$ where  
the infinitesimally low-level equilibrium value $x_\ell \ll 1$ lies.
Separating the basins of attraction of $x_{\rm h}$ and $x_\ell$ is
a critical point $x_{\textrm D}$, where the phase flow diverges.
At $x = x_{\rm D}$ two optimal water strategies $u^*$ with identical optimal
payoff coexist: the low-intensity one for $x \to x_{\textrm D}^+ $ leads to intensified
terrorist strength so that $x \rightarrow x_{\rm h}$,
whereas the high-intensity one for $x \to x_{\textrm D}^- $
results in near eradication of terrorist activity with $x \rightarrow x_\ell$.
The vertical dotted line denoted by $x_\mathrm{S}$ indicates
the critical terrorist strength below which optimal fire strategies
$v^* \to 0$, suggesting that the optimal way to counter initially weak  
terrorist groups is to use only water strategies.
Taken and modified from Ref.\,\cite{CAU08}.}
\label{FIG:CAU08}
\end{figure}

Fig.\,\ref{FIG:CAU08}(a) shows that as $x$ increases, the optimal intensity
of fire strategies $v^*$ also increases, signifying that stronger terrorist groups require 
more vigorous intervention, whereas the optimal water intervention $u^*$ remains
approximately constant. For $x$ large enough however, the dynamics of the system
drives terrorist strength towards a finite, non zero Nash equilibrium $ x \to x_{\mathrm h}$.
Since payoff functions aim to minimize terrorist
activities but also government expenses,  eradication of the extremist groups does
not necessarily yield the optimal utility.  
Fire tactics are thus best suited when the initial strength of the terrorist organization 
is larger than its equilibrium, $ x > x_\mathrm{h}$, since they 
will lead to a partial weakening of the extremist group.

At small values of $x$, as shown in the magnified view in Fig.\,\ref{FIG:CAU08}(b),
a switch point $x_{\textrm S}$ exists, whereby 
the optimal fire strategy $v^* \to 0$ as $x \to x_{\textrm S}$ from above. 
For $x < x_{\textrm S}$, the constraint $v \ge 0$ becomes activated, keeping $v^* = 0$.
This result implies that when terrorists are not very resourceful, water strategies are the sole optimal way
to completely eliminate terrorism. 
At low values of $x$, a critical Dechert Nishimura Skiba (DNS) point 
$x_\mathrm{D}$ \cite{SKI78,DEC83} arises. 
In dynamical optimal control theory DNS points represent loci 
in state space where the same initial condition leads to 
multiple optimal trajectories, each converging to distinct long-term steady states, 
all of which are optimal.  Fig.\,\ref{FIG:CAU08}(b) shows the 
two distinct optimal trajectories $u^*$ originating from $x = x_\mathrm{D}$.
The lower-intensity strategy $u^*$, for $x \to x_{D}^+$, 
 leads to the $x_{\rm h}$ equilibrium, 
along the dynamics described in Fig.\,\ref{FIG:CAU08}(a).
The higher-intensity $u^*$ for $x \to x_{D}^-$ drives the dynamics 
towards a distinct and infinitesimally low Nash equilibrium $x_\ell \ll 1$,
representing near-elimination of terrorist activities.
Variants of the fire and water model 
here described, as well as other optimal control methods, have also been 
developed to study how to best design
counter-terror tactics \cite{KRE09, SEI16, BAY19, AZH22}. 

Citizen approval of anti-terrorism operations 
is introduced to the model defined in \cite{CAU09} 
through an additional state variable $y(t)$ that represents support from the general public.
This quantity increases as terrorist organizations grow their resources $x$, 
but decreases under aggressive counter-terrorism operations 

\begin{equation}
   \frac{\mathrm{d} y}{\mathrm{d}t}  = r_y x - \omega u_2^2 + \zeta (y_\mathrm{max} - y),
\end{equation}

\noindent
where $r_y$ is rate of public support,
$\omega$ describes antagonistic effects due to overaggressive counter-terrorism initiatives, 
and $\zeta$ specifies the rate of $y$ converging to a maximum $y_\mathrm{max}$ under neutral
conditions. Higher public approval is also modeled to increase the effectiveness of 
counter-terrorism operations, as it may lead to larger budgetary means to combat
extremism, or even in active participation in seeking fugitives or reporting
suspicious activities.  
A $y$-dependence is thus 
introduced to the harvest function $h(y,u_1, u_2)$, such that $\partial h / \partial y > 0$
for $x, u_1, u_2 > 0$.
Multiple equilibria are found depending on parameter choices. In one case, the optimal scenario is the
counter-terrorism authority enjoying public support and effectively suppressing terrorism; 
in the opposite case negative public opinion hinders anti-terrorism efforts and allows extremists to thrive. 
Finally, a specific set of parameters yields bistable solutions depending on initial conditions:
here, strong public support at the onset of counter-terrorism 
operations is shown to be crucial for terrorist activity to be effectively contained. 
Sensitive regions in phase space
are also identified whereby small parameter perturbations may strongly affect
the final equilibrium configuration: such volatility may represent both a risk and an opportunity
for counter-terrorism agencies as they develop intervention strategies.

A three-way interaction model that includes counter-terrorism efforts, public opinion, and a community that harbors terrorists
is presented in \cite{DRA14}.  The harboring community is assumed
to contribute to anti-terrorism efforts, as long as the scale of these
operations is limited. However, once governmental involvement 
exceeds a given threshold, the harboring 
community becomes antagonistic: costs are increased and
the effectiveness of counter-terrorism efforts is reduced. 
Public opinion on the other hand is assumed to be supportive of anti-terrorism programs that
decrease attack risks; these programs may be costlier. As a result, counter-terrorism agencies must weigh
whether to increase the $u_2$ intensity of their operations to garner public support, 
or risk alienating the harboring community, increasing related costs.
It is found that by applying legal constraints to counter-terror efforts, effectively
imposing bounds to the acceptable values of $u_2$, public opinion and the harboring community may both be steered towards 
being more supportive of governmental intervention. Similar issues
were explored in Stackelberg games where limits were posed on interrogation methods
and other tactics to extract information from detainees \cite{BAC09}. 

The study of transnational terrorism utilizes variants of multi-person games \cite{OSB94, TAD13}, 
where competitions may arise among multiple nations, or among nations and terrorist organizations.
A nation may cooperate with other countries to intervene against 
terror groups, or opt to avoid the associated high costs and to instead rely on actions taken by other countries, 
``passing the buck" and avoiding attention. 
A nation may even support, covertly or not, the operations of a terrorist organization against other states. 
These scenarios are studied via game-theoretic models that incorporate political and economic components of terrorism \cite{HAU21}.
Specifically, three labor stocks of a terrorist organization are considered: the ideologues, volunteers deeply committed to the terrorist organization’s political beliefs; 
the criminal mercenaries, motivated primarily by economic opportunism; 
the captive participants, who have been threatened to serve the organization.  
Terrorist organizations are assumed to draw on the three labor stocks as they conduct attacks; 
specific sponsor parties can help replenish some of them, depending on the type of attack.
For their part, governments need to devise appropriate interventions to reduce the three stocks, 
which comes with political, economic, and human costs. If costs are too steep,
no action is taken, under the expectation that other  countries will. The model also allows
for governments to become a stock-replenishing sponsor, if political or economic gains can be projected. 
Results show that different circumstances lead to very diverse dynamics, 
and that no global optimal strategy exists. The multi-faceted aspect of the problem
and the many branches of possibilities,  in a way, confirm that there are
many challenges in eradicating, or reducing, terrorist endeavors. 

Game theoretic models have also been applied to study terrorism from different perspectives.
For example, \cite{FEN16,REZ17} optimized patrol schedules to protect crucial government assets from 
terrorist attacks, such as oil pipelines and chemical plants, under the constraint of limited resources; 
in the same vein, another study considered a leader-follower game where the state (the leader)
installs facilities and a terrorist group (the follower) attacks metropolitan areas  \cite{BER07}. The game is played out so
that losses due to terrorist attacks are minimized by the strategic placement of these facilities. 
How to best allocate resources to defend sites at risk is also discussed in \cite{GOL09} where
the distinction is made between damage due to chance events and targeted action by ill-intentioned parties. 
In the former case, the optimal policy is to invest resources on target high-priority sites, in the latter
to spread resources and protect even the most vulnerable areas. 
Other defense strategies are presented in \cite{HAU11a, HAU11b, HAU17}. 
The synergy between terrorists and the active or passive support of one or more state sponsors
as they form a coalition to attack a third party target has also been studied \cite{MCC09}. Here, the interplay between players changes in time: 
the state sponsor may use the terrorist group as a way of covertly promoting 
hostile acts towards the third party, but the latter may increasingly become aware
of manipulations and retaliate. 
Brinkmanship game theoretical models have also been proposed: here one or both parties push dangerous actions, 
on the brink of disaster, to get the most advantageous outcome.  
The challenge is to issue threats that must be sufficiently unpleasant to deter terrorists, but not so repugnant 
as to not be carried out; these models thus carry a credibility constraint \cite{MEL09}.
The effects of counter-terrorism policies have been studied in two strategic cases \cite{ARC09}:
when government intervention is defensive, and a corollary to terrorist action,
and when it is a proactive substitute for it.
Negative responses may arise in both cases: erosion of terrorist support, if
attacks cause too much damage; backlash against the government, if its response is too strong.
It is found that large-scale attacks will occur if government reactions produce a strong enough
backlash. Game-theoretic network centrality, which identifies the most influential nodes within a cooperative game, 
has proven to be useful in identifying key members of terrorist networks \cite{MIC15}.
Search games have also been proposed as methods to hunt key radical operatives: 
terrorists attempt to maximize the time to search terrorist networks, while governments
seek to minimize it \cite{FOK13}. The scenario of 
former terrorist organizations seeking a peaceful resolution to ongoing
conflict has also been analyzed \cite{DEM05}.
Here, a game of learning is used to describe negotiations
between the two parties, whereby governmental agencies infer
the willingness and the ability of terrorists to commit to the peace process
by observing their actions.  Evaluations are then used by the 
anti-terrorism agency to decide whether to keep pursuing negotiations 
or abandon them. Finally, since transnational terrorist activities can affect policy making in different countries,
sequential games have been introduced involving multiple governmental players
so that a country may choose between a preemptive or a defensive counter-terrorism
strategy in response to the outcome of another country's strategy. The two 
may also rationally evaluate whether it is beneficial to forge an anti-terrorism coalition 
or engage individually \cite{SAN06,DEO18}.

\section{Terrorist events as self-exciting processes 
\label{sec:self-exciting}}

One major question in the study of terrorism is whether the risk of future attacks can be estimated, drawing
on information gathered from past events and using mathematical modeling, 
risk management or machine learning methods
\cite{SUB13a, DIN17, BUN06, FEI10, SUB13b, MAN13, HAS18}. It is well known that terrorist activity is not entirely random and that 
a prior attack can temporarily increase the likelihood of another occurrence in its geographical vicinity  \cite{LEW12,TEN16}.
First attempts in this direction are found in early studies of terrorism where various time-series analyses
were used to study the frequency of terrorist attacks \cite{HOL87, END93, END00, END02, BAR03, NAG05, END06}, 
and to explain both the temporal clustering of events and the long quiescent periods in between \cite{MID80, DUG05,LAF09,LAF10}.

Mathematically, the phenomenon of past events acting as catalysts for future ones 
can be described as a self-exciting temporal point process, 
also known as the Hawkes process \cite{HAW71,HAW74} which was first introduced to study
contagious phenomena and to predict earthquake aftershocks \cite{OGA88,HEL03}.
Given that social behaviors are often of a contagious nature and transmit through
human interactions, a wide range of social activities have been studied as
self-exciting phenomena. These include committing crimes \cite{SHO08, SHO09, MOH11}, the sparking of protests \cite{
GOV81, MYE00,  DAV13, BON18},
the erupting of political disorder and violence \cite{MID78, CAM21, BAH24}, the occurrence of political turnovers \cite{SCH00} and
military coups \cite{LI75}, the clustering of suicides \cite{GOU90}, and even the exchange of email messages \cite{FOX16}. 
Within the context of terrorism, self-excitation markers have been found in airplane hijacking incidents
during the 1980s \cite{HOL86}, and in insurgent activities in the early 2000s \cite{TEL06, TOW08, BEH12, BRA12}.
The canonical form of a discrete Hawkes process can be expressed as
\cite{ LEW12, HAW71, HAW74, OGA88, SHO08, MOH11, HOL86, SHO09}

\begin{equation}
   \lambda (t) = \mu (t) + \sum_{i: t_i < t} \nu (t - t_i), 
   \label{EQ:SELF_EXCITING}
\end{equation}

\noindent
where $\lambda (t) > 0$ specifies the rate of a terrorist attack at
time $t$, $\mu (t) > 0$ is the background rate, and the response function $\nu(t -t_i)> 0$ 
defines the elevated rate at time $t$ in response to an event at a prior time $t_i$. The quantity 
$\lambda(t) - \mu(t)$ is the self-excitation contribution to the terrorist attack rate.
Typically $\nu(t-t_i)$ is a decreasing
function of its argument so that the enhanced likelihood decays as 
$t_i$ moves into a more distant past. While other forms of decay functions
have been adopted in the literature \cite{POR12,WHI13,KHR16}, a common choice is
the exponential form \cite{LEW12,TEN16,SHO08,MOH11,SHO09,KHR16,CLA18}

\begin{equation}
   \nu ( s ) = k_0 \omega \exp ( - \omega s ),
   \label{EQ:DECAY_FUNCTION}
\end{equation}

\noindent
 where $\omega^{-1} > 0$ is the time scale over which self-exciting effects
dissipate, and $k_0 = \int_0^\infty \nu (s) \mathrm{d} s$ 
represents the increased attack rate immediately after a prior one.
Eqs.\,\eqref{EQ:SELF_EXCITING} and \eqref{EQ:DECAY_FUNCTION} with a constant 
$\mu (t) \equiv \mu_0$ are the most basic version of a self-exciting model for
terrorist attacks, consisting of only three parameters: $\mu_0$, $k_0$, and $\omega$.
For a given time series data $0 \le t_i \le T$ for $i = 1, 2, 3, ..., n$, parameters 
can be fitted through the methods of maximum likelihood estimation (MLE), which finds a set of
parameter values that maximizes the following log-likelihood function $\mathcal L$ \cite{FIS22, RUB72,OZA79}

\begin{equation}
   \log {\mathcal L} = \sum_{i=1}^n \log \left( \lambda ( t_i ) \right) - \int_0^{T} \lambda (t) \mathrm{d} t
   \label{EQ:LOG_LIKELIHOOD}.
\end{equation}

\noindent
White et al. interpret the three parameters
that appear in Eq.\,\eqref{EQ:SELF_EXCITING}
respectively as the ``intrinsic risk'' ($\mu_0$), ``volatility'' ($k_0$), 
and ``resilience'' ($\omega^{-1}$) of terrorist activities within a country  \cite{WHI13}.
The same authors apply the self-exciting model in Eqs.\,\eqref{EQ:SELF_EXCITING}--\eqref{EQ:DECAY_FUNCTION}
to terrorist events between 2000 and 2010 in three southeast Asian countries -- Indonesia, Philippines, and Thailand -- finding 
significant variance among them,  consistent with their respective sociopolitical contexts. 
This study demonstrates that the simple self-exciting 
model can be useful in quantitatively assessing 
terrorist threats in different geopolitical regions. 
Regional variations were also found in the underlying 
mechanisms of daily death tolls in four distinct Iraqi districts, which were analyzed between
2003 and 2007 
\cite{LEW12}, and between Syria and England, which were analyzed between 2010 to 2014 \cite{FUJ17}.
Najaf, the relatively more quiescent Iraqi region among the four districts examined in \cite{LEW12}, 
displays the lowest intensity and shortest
decay time of self-excitation.  Syria exhibits highly 
clustered, self-excited civilian deaths, while England mainly displays
only a background death rate \cite{FUJ17}.
 Regional variations in the self-exciting parameters 
$\mu_0$, $k_0$ and $\omega$ due to local socio-politics imply that these parameters
may also change temporally. This is shown in 
\cite{TEN16} where the dynamics of more than 5,000 
improvised explosive device (IED) attacks by the Provisional Irish Republican Army (PIRA)
in Northern Ireland was analyzed between 1970 to 1998.
This period was divided
into five phases, each defined by major events within PIRA's history, and 
Eqs.\,\eqref{EQ:SELF_EXCITING}--\eqref{EQ:DECAY_FUNCTION} were applied to each. 
Significant changes in model parameters were found across the five time intervals: 
for example, when PIRA fractured into smaller cells, 
increased background rates, decreased volatility, and 
faster dissipation of the self-exciting effects were observed.

 \begin{figure}[t]

\includegraphics[width=4.4in]{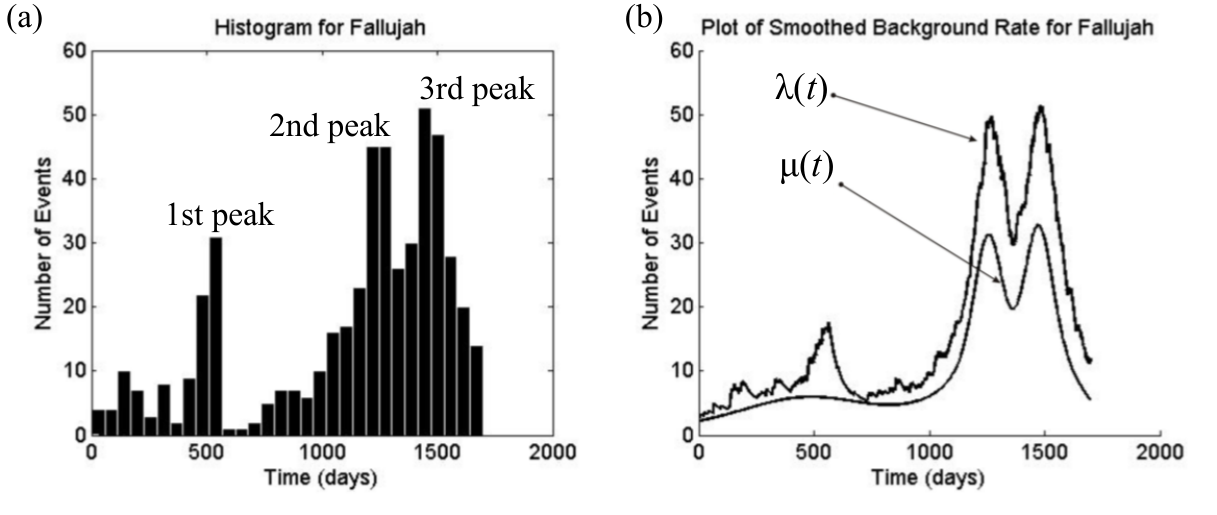}
\caption{Panel (a) shows the histogram of number of fatalities per day
in Fallujah, Iraq from March 20$^{th}$ 2003 to December 31$^{st}$ 2007.
Note the emergence of three peaks. 
Panel (b) shows the total $\lambda (t)$, 
and time-dependent background $\mu (t)$, rates fitted from the data in panel (a).
Their difference, $\lambda(t) - \mu(t)$, represents the self-excitation 
component of the terrorist attack rate. 
It is interesting to note that the first peak, around day 500,  
is mostly attributed to self-excitation, whereas the second and third
peaks, around days 1250 and 1500 respectively, 
are driven primarily by increases in the background rate.
Taken and modified from Ref.\,\cite{LEW12}.}
\label{FIG:LEW12}
\end{figure}

While the sparsity of regional terrorism records often hinders the applicability 
of overly sophisticated models \cite{POR12,WHI13}, 
the basic Hawkes process still allows for
the inclusion of time dependence in its three fundamental 
parameters, particularly in the background rate $\mu (t)$, which 
should naturally reflect local socio-political scenarios.  
Simple time-dependent functions for $\mu (t)$ include step functions, 
ramp functions, and heat kernels \cite{LEW12,KHR16,JOH18}. 
An example of the self-excitation model 
using a time-dependent $\mu (t)$ is displayed in Fig.\,\ref{FIG:LEW12}, 
fitted for the daily death tolls in Fallujah, Iraq from March 20$^{th}$ 2003 to December 31$^{st}$ 2007, 
showing the respective contributions from the background
rate $\mu (t)$, and the self-exciting process $\lambda (t) - \mu (t)$.
Lewis et al. \cite{LEW12}, Khraibani and Khraibani \cite{KHR16}, and Johnson et al. \cite{JOH18} 
observe that given sufficient data, models with a constant $\mu (t) = \mu_0$ are either inadequate or consistently
outperformed by models with time-dependent $\mu (t)$. For more flexible time
dependence, a nonparametric method of Maximum Penalized Likelihood Estimation (MPLE)
is proposed in \cite{LEW12b}. Without assuming any specific functional form for
$\mu (t)$ and $\nu (t)$, the method finds the best fitting functions via a
log-likelihood function modified from Eq.\,\eqref{EQ:LOG_LIKELIHOOD}

\begin{equation}
   \log {\cal  L} = \sum_{i=1}^n \log \left( \lambda ( t_i ) \right) - \int_0^{T} \lambda (t) \mathrm{d} t
            - \alpha_1 R(\mu) - \alpha_2 R(\nu),
   \label{EQ:LOG_LIKELIHOOD2}
\end{equation}

\noindent
where $R$ is a roughness penalty function (e.g., the $L^2$-norm of the first-order
derivative of $\mu(t)$ and $\nu(t)$) which ensures certain regularities of the time-dependent
functions $\mu(t)$ and $\nu(t)$, and $\alpha_{1,2}$ are the respective weights \cite{EGG01}. 
The basic self-exciting process in Eqs.\,\eqref{EQ:SELF_EXCITING}--\eqref{EQ:DECAY_FUNCTION}
can also be extended to include mutual-excitation,
whereby a distinct set of occurrences occurring at times $\{ \tau_j \}$ 
further influence $\lambda(t)$. 
A possible representation is  \cite{TEN16}

\begin{equation}
   \lambda (t) = \mu (t) + \sum_{i: t_i < t} \nu (t - t_i)
      + \sum_{j: \tau_j < t} \eta (t - \tau_j), 
   \label{EQ:SELF_EXCITING2}
\end{equation}

\noindent
where the $\{ \tau_j \}$ timings exert their influence on 
$\lambda(t)$ through the cross-response function $\eta(s)$. 
The original study on the distribution of IED attacks by PIRA \cite{TEN16}
was revisited using Eq.\,\eqref{EQ:SELF_EXCITING2}  to examine whether attacks in different North Ireland 
counties and its capital Belfast could influence each other. From the analysis it
was concluded that geographic mutual-excitation was far less influential than self-excitation. 
A similar conclusion was reached in a parallel study of
terrorist attacks in Iraq between 2003 and 2010. \cite{CLA18}. Here, spatial diffusion was included in the background rate $\mu(t)$, 
but the resulting analysis failed to capture the regional spread of terrorist activities.
In addition to different geographic regions, the mutual-excitation model 
in Eq.\,\eqref{EQ:SELF_EXCITING2} can also be used to examine interactions
between different actors. For example, the 1970-1988 IDE attacks by PIRA were further 
analyzed to verify whether the corresponding $\lambda(t)$ was affected by the actions of the
British Security Force (BSF) \cite{TEN16}.  One interesting finding was that PIRA's activities 
intensified much more as a reaction to Catholic civilian deaths caused by BSF, 
rather than to direct BSF killings of PIRA members.  Such cross examinations may provide
valuable insight for policy makers in devising guidelines for counter-terrorism operations.
Further work combined self-exciting processes with the so-called ``hurdle" model 
to analyze the daily number of terrorist attacks in Indonesia between 1994 and 2007 \cite{POR12}.
The hurdle model describes the frequency-position and the severity-height
of events as two separate stochastic processes \cite{MUL86,HEI94,CLA12}. In the 
work of \cite{POR12} the dates and number of terrorist attacks per day are 
chosen as the frequency and severity of the hurdles, respectively. 
Using a Riemann zeta distribution for the 
severity component, the authors found that incorporating self-exciting processes in the frequency 
component improves the fitting of the hurdle model to the data, and confirmed the
hypothesis of a short-term increase of terrorism risk following an attack.

As for why terrorist activities exhibit self-exciting characteristics,
some believe the reason is economic, as it may be more efficient in terms of 
costs and benefits to strike as many times as possible before any window of 
opportunity closes \cite{END05, BER07}.
Others point to copy-cat behaviors, as terrorists in the same organization
or like-minded extremists tend to learn from each other. 
A positive-feedback may thus be generated through imitation, so that a single act can 
give rise to a cascade of events \cite{CLA12}. 
The relationship between the harboring of radical beliefs and the execution
of terrorist attacks may be also addressed using hidden Markov models, where radical opinions can be 
regarded as a hidden layer of latent states, from which terrorist attacks manifest as observables \cite{PET12,RAG13}.
Regardless of the underlying reason, counter-terrorism
intelligence may exploit the non-uniform likelihood of terrorist attacks to strategically
time possible interventions and allocation of resources. 

\section{Terrorism through the lens of data}
\label{data}

A primary goal of mathematical modeling is to offer a quantitative framework 
of the possible mechanisms that drive observed phenomena, and to
facilitate the forecasting of future events. 
Being able to fit models and theories to real-world data represents a significant step forward.  
Due to the covert nature of terrorist operations, and the relative scarcity of accessible, collated data, 
this task has been particularly challenging in terrorism-related research. 
The recent development of data-mining techniques however, has
allowed significant advancements. For example, text-mining and lexical analysis
have made it feasible to retrieve meaningful text and numerical data from various sources 
of media reports,  leading to the compilation of large, freely available datasets. 
Cluster analysis is an immediate and powerful application of such datasets. Here, 
similar data points are grouped together based on their characteristics so that 
patterns, relationships, and other commonalities can be identified. 
In particular, datasets containing Geographic Information Systems (GIS) 
information can help identify 
spatial and temporal clusters of terrorist events, 
following the theory of terrorist activities as self-exciting processes 
described in Sec.\,\ref{sec:self-exciting}. 
The study of spatiotemporal clustering, also known as near-repeat analysis, 
helps assess the recurrence risk of terrorist attacks \cite{TEL06, TOW08, BEH12, BRA12,  KHR16, CLA18, CHU19b}. 
Network analysis may impart further refinements, for example by helping 
identify subgroups of numeric and categoric data. One can thus uncover similarities 
among terrorist events or the {\em modus operandi\/} of terrorist groups, 
such as method of attack, number of victims, proximity to specific locations
(schools, churches, government offices).  
The resulting patterns and structures may guide the 
application of unsupervised machine-learning techniques, 
leading to the construction of predictive models. 

Efforts to catalogue and maintain records of terrorist events began in the 1970s. 
One of the pioneering projects started in 1972 and was led by Brian Jenkins of the RAND Corporation, 
with the support of the Department of State and the Defense Advanced 
Research Projects Agency (DARPA) of the United States.  The goal was to 
compile and maintain a chronology of incidents of international terrorism \cite{JEN75}. 
Around the same time, Edward Mickolus of the United States 
Central Intelligence Agency (CIA) began the ``International Terrorism: 
Attributes of Terrorist Events'' (ITERATE) project to keep track of 
the activities of transnational terrorist organizations \cite{MIC76}. 
Researchers from the Pinkerton Global Intelligence Service (PGIS), 
a private international detective agency, also aimed to create a list 
of terrorist incidents as collected from domestic and foreign newspapers, wire services, 
and government reports \cite{LAF10}. 

By the early 1980s, these datasets had amassed thousands of 
entries detailing terrorist events from 1968 through 1980 \cite{MIC80, FOW81}. 
The spread of the internet in the 1990s, and the advent of digital publications, 
granted wider accessibility to journalistic and other written reports of terrorist
incidents.  In addition, the development of data-mining techniques
 helped accelerate
the process of data acquisition.  As a graduate student at the University of Bergen in 1992, Jan Oskar Engene 
began constructing a dataset called  ``Terrorism in Western Europe: 
Events Data'' (TWEED). By the time the project was completed, 
TWEED had gathered more than 11,000 records of terrorist 
incidents in Western Europe between 1950 and 2004
\cite{ENG07}.  The RAND list grew from a few thousand entries in the 1980s to over 40,000 
before updating stopped in 2009.  

The PGIS list, which collected data from 1970 through 1997, 
was transferred in 2001 to the National Consortium for the Study of Terrorism 
and Responses to Terrorism (START) at the University of Maryland, College Park \cite{LAF07}. 
At the time, it had logged in more than 67,000 incidents worldwide and was the most comprehensive
of all existing archives.  In 2006, START received funding from the National Institute of Justice
to build the Global Terrorism Database (GTD) designed 
to validate and computerize the PGIS data, and to extend it beyond 1997. 
Due to changes in the managing entities and methodologies employed, the GTD exhibits some discontinuity 
for data collected before and after 1997 \cite{END11}.  
It is also well-known that the original 1993 data was almost 
entirely lost as files were transported between offices. 
Despite all these challenges, the GTD has emerged as one of the 
most comprehensive and influential datasets for terrorism research. 
At this time of writing, it has compiled more than 200,000 records of terrorist events from 1970 through 2020. 
Also widely used is ITERATE, now maintained by Vinyard Software, a company presided by Mickolus. 
In its current form, ITERATE has a narrower scope than the GTD, focusing on transnational terrorist 
attacks against civilian targets, excluding those occurring on US soil and against military targets. 
It also records attributes such as the fate of perpetrators,  negotiation details in hostage situations, 
and flight information in airplane highjackings.
Aside from the more prominent ones listed here, 
other datasets of terrorist incidents are available, 
some with a more specialized scope, others focusing on specific geographic areas. 
New data is continuously retrieved, verified and added to all actively maintained databases 
\cite{BOW17,BOW20,BOW21}.

In Sec.\,\ref{sec:self-exciting}, we reviewed works that model
terrorist events as self-exciting processes, a direct
outcome of which is the spatiotemporal clustering of events. 
One of the first studies to cluster groups of countries (out of a total of $n$ possible countries)
based on the occurrence of terrorist events ITERATE \cite{BRA07}. In this study, each country $i$
with is labeled a ``hot-spot"
depending on its Getis and Ord autocorrelation index $G_i^*$ defined as \cite{ORD95} 

\begin{equation}
   G_i^* = \frac{\sum_{j \neq i}^{n}  w_{ij} \left( x_j - \bar{x} \right)}
                {\hat{\sigma}_x \sqrt{
                         \left[ n \sum^n _{j \neq i}  w_{ij}^2 - \left( \sum^n _{ j \neq i} w_{ij} \right)^2 \right]
                         \Big{/} \left( n - 1 \right) } }, \qquad 1 \leq i \leq n.
                          \label{eq:GIndex}
\end{equation}

\noindent
Here, $x_j$ is the number of terrorist events occurring in country $j$ 
from 1975 to 1997, $\bar{x}$
and $\hat{\sigma}_x$ are the mean and variance associated to the $x_j$'s, respectively. 
The quantity $w_{ij}$ is an element of a binary contiguity matrix 
where $w_{ij} = 1$ if countries are neighbors according to given
definitions (involving land and sea contiguity), and $w_{ij} = 0$ otherwise. 
Finally, $n$ is the number of countries examined, here set at $n=112$.  
As defined in Eq.\,\eqref{eq:GIndex}, positive $G_i^*$ values are associated to 
countries for which local averages are higher than global averages. 
Country $i$ will be considered part of a hot-spot if its $G_i^*$ is positive and statistically significant,
according to specific tests detailed in \cite{BRA07}.  Although other localized spatial statistics could have been chosen
(such as local Moran’s $I$, local Geary’s $C$, and the Getis and Ord $G_i$  
\cite{CRA00, OLO02, GLE02, CHA05}),  Eq.\,\eqref{eq:GIndex} is considered a superior metric as its
mathematical definition is closest to the intuitive concept of a hot-spot. 
Using three different definitions of contiguity, 
the authors find that the designation of geographical hot-spots is consistent and robust. 
The temporal correlation is investigated by dividing the available 23 years of data
into batches of three-year windows, showing elevated risks of terrorist attacks following previous incidents
and revealing that hot-spots persist over lengthy periods.

Spatio-temporal correlation also emerged from analyzing 
terrorist attacks attrib\-uted to the El Salvador-based guerrilla group Farabundo Martí National Liberation Front (FMLN) 
(more than 3,000 events between 1980 and 1992) and 
the separatist Euskadi Ta Askatasuna (ETA) movement in the Basque country, Spain
(almost 2,000 events between 1970 and 2007).
Both organizations, although different in terms of history and motive, were found to 
preferentially employ bombings and
non-lethal attacks, and to target national or provincial capitals \cite{BEH12}. To identify
localized bursts of near-repeat terrorist activities, dubbed ``microcycles'' of violence, 
the authors employed the Knox test, which was first developed in 
epidemiology to detect non-random clusters of diseases \cite{KNO64, KNO64b}. 
This test examines whether the concentration of data points 
within a distance and time threshold exceeds random chances 
by comparing observations with samples generated by random permutations
of the time and location of the actual observations. 
For a fixed spatial area and temporal window, 
the ratio of the number of paired, consecutive events from observations 
over the number of pairs from randomized data is called the Knox ratio.
Values greater than 1 suggest spatial clustering, values 
less than 1 indicate random distributions. 

\begin{figure}[t]
\includegraphics[width=4.7in]{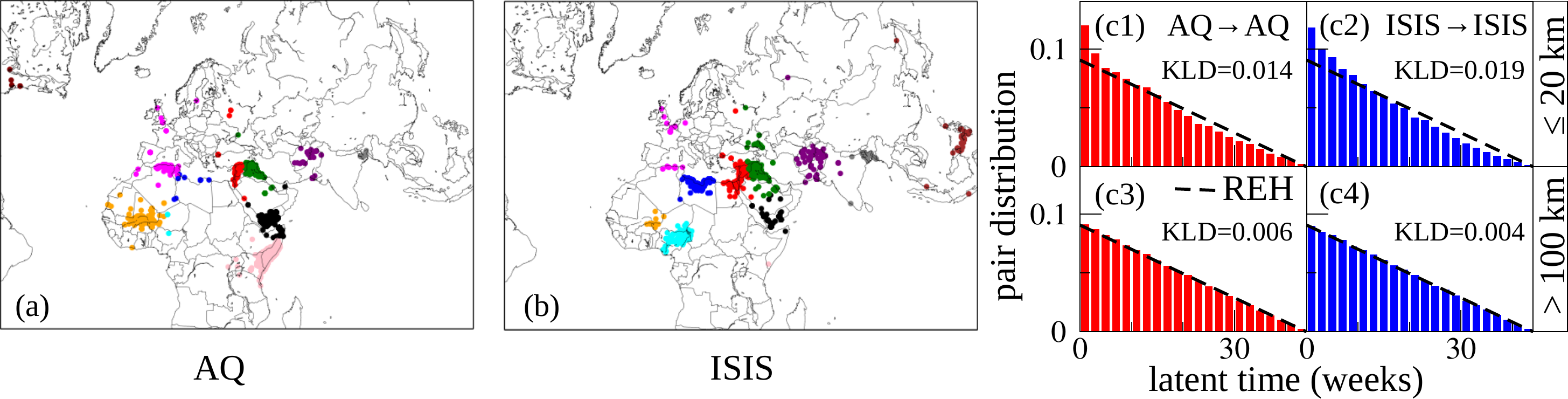}
\caption{Clustering and near-repeat patterns of terrorist events
perpetrated by Al- Qaeda (AQ) and ISIS (IS) between  2001 and 2017 using the GTD. 
Panels (a) and (b) highlight the twelve geographic clusters where 
AQ and IS attacks co-localize. Panels (c1) -- (c4) show 
near-repeat attack statistics for AQ and IS compared
to the REH line given by $P_{d}(t)$  in Eq.\,\eqref{eq:REH} and plotted as a dashed line.
The bar graphs show the distribution of time intervals $t$ between attacks, with
each bar representing two weeks. Panels (c1) and (c2) consider near-repeat attacks separated
at the most by $d=20$ km; in panels (c3) and (c4) instead $d=100$ km. 
Panels (c1) and (c2) show that the likelihood of a near-repeat 
attack within $d=20$ km is higher than the REH for the first 4 (AQ-AQ) and 
10 (IS-IS) weeks after a given attack. 
Instead, the likelihood of a near-repeat attack is the same as the REH
for attacks separated by more than $d=100$ km. The KLD 
value is a measure of the overall deviation between the bar graph and the REH
and helps quantify near-repeat attack tendency.  The largest near repeat likelihood is
for IS-IS attacks.  Taken and modified from Ref.\,\cite{CHU19b}.}
\label{FIG:near_repeat}
\end{figure}

Following the September 11 attack by al-Qaeda (AQ) in 2001 and 
the subsequent 2003 US-led invasion of Iraq, transnational terrorist attacks
notably intensified, culminating with the rise of the Islamic State (IS), 
which at its peak controlled more than 100,000 $\mathrm{km}^2$ of territories 
in Northern Iraq and Eastern Syria. It also created numerous remote branches 
in other Islamic countries, and created ties to other local insurgent groups.  
To investigate the interplay of terrorist activities attributable to
AQ and IS,  the two dominant transnational organizations in the early part of the $21^{\textrm {st}}$ century, 
Chuang et al. \cite{CHU19b} analyzed GTD entries of their respective
attacks between 2001 and 2017. 
In contrast to the more coarse ITERATE dataset where geographical information
is at the country level, the GTD dataset reports longitude and latitude coordinates 
of each terrorist incident, providing a more refined spatial resolution.  
Using the $k$-means algorithm \cite{MAC67}, 
twelve geographic clusters where AQ and IS attacks co-localized were identified, 
as plotted in  Fig.\,\ref{FIG:near_repeat}. 
The propensity of AQ and IS to conduct repeat attacks 
within given geographical areas was then investigated for each cluster, 
based on the theory of terrorist events as self-exciting processes, 
described in Sec.\,\ref{sec:self-exciting}. 
Time intervals $t$ separating pairs of attacks within
a fixed time window $w$ and separated by a maximum distance $d$ 
were evaluated and compared to the theoretical distribution generated from the random-event hypothesis (REH)

\begin{equation}
   P_{d} ( t ) = 2 \frac{w - t}{w ( w - 1 )}. \label{eq:REH}
\end{equation}

\noindent
The discrete quantities $t, w$ are measured in weeks or months and since
$0 < t < w$,  $\sum_t P_{ d} (t) = 1$. The REH is built on the hypothesis that 
if attacks occur randomly, then the likelihood of two events being separated
by time $t$ is proportional to the number of pairs separated by $t < w$ that can be
accommodated within the time window $w$. 
A higher percentage of the observed distribution at short-time intervals
relative to the REH is indicative of near-repeat attacks. 
This is the case for near AQ-AQ repeat and near IS-IS repeat 
attacks as shown in Fig.\,\ref{FIG:near_repeat}(c). 

Figs.\,\ref{FIG:near_repeat}\,(c1) and (c2) reveal that for $d= 20$ km
near-repeat tendencies, measured as discrepancies of observed data from the 
 REH in Eq.\,\eqref{eq:REH}, emerge for both AQ-AQ and IS-IS attacks, and over time periods
 of 4 and 10 weeks, respectively. 
In contrast, Figs.\,\ref{FIG:near_repeat}\,(c3) and (c4) show that terrorist
events separated by at the most $d = 100$ km do not display near-repeat
tendencies, are randomly distributed and follow the REH in Eq.\,\eqref{eq:REH}.
To quantify near-repeat tendencies, 
the Kullback–Leibler divergence (KLD) was used \cite{KUL51}, so that
deviations between the data derived from the GTD and the REH line could be quantified.
The KLD values reveal a stronger near-repeat tendency for IS-IS attacks over AQ-AQ attacks. 
Near-response attacks were also investigated, whereby
an attack by a given group (AQ, IS or local insurgents) elicits a response from a different one. 
Patterns were seen to 
depend on the adversarial, neutral, or collaborative relationship between groups. 
When in conflict, local insurgents respond quickly to attacks by either AQ or IS, 
while the latter delay their response, leading to an asymmetric dynamic. 
When neutral or allied, attacks by one group enhance the response likelihood of the other, regardless of hierarchy,  
group relationships, such as cooperation, rhetorical or violent rivalry, or peaceful coexistence.
The consequences of government intervention, spillover effects (terrorist activities  spreading from one region to neighboring ones), 
outbidding (escalation of events due to rivalry between terrorist groups) were also discussed in mathematical terms.

As the above results highlight, cooperation and conflict among terrorist groups, and how they interface with local 
insurgent groups or the government,
are key determinants of terrorist activity, and may lead to escalation or restraint. These complex relationships 
however, remain  relatively understudied \cite{POL24}.  
A set of 260 AQ and IS affiliates and factions
was analyzed through time-series of triangles and triadic relationships
to verify whether the postulates of social balance theory are followed \cite{BEL20}. 
About 2,700 relationships between groups were compiled 
from primary sources, including propaganda, intercepted documents, declassified intelligence reports
and social media posts, as well as from books and newspapers.  These relationships were
given one of five ratings between ``strongly cooperative" and ``strongly adversarial."
According to social balance theory,  a triangle of cooperative factions is stable, and so is 
one where one faction is in an adversarial relationship with other two, who instead 
cooperate with each other. Within the AQ and IS ecosystem, the authors show that 
the long-term stability of balanced triangles is not guaranteed, and that strategically
disrupting allegiances among factions can help destabilize terrorist networks.  
Other studies address the geographical distribution and possible spatio-temporal clustering
of terrorist events in specific regions or countries such as Jerusalem, Turkey, Egypt, Pakistan, Thailand, 
\cite{PER20, HAC21,YOU23, IMR23, CHI23} or as perpetrated by specific groups such as the Taliban \cite{RIE21}.
The observed clustering of terrorist events leads to the further question
of whether any possible spatial determinants of risk can be identified.  
It was found that terrorist attacks in Israel typically clustered near the operational bases of the terrorists 
or near disputed borders \cite{BER07b}. On a more local scale, an analysis of more than 1,000 terror attacks occurring
in Istanbul between 2008 and 2012 found that religious, educational, health 
facilities and government buildings represent high-risk locations
and that the terrorist group's ideology played a part in selecting the location of the attack \cite{ONA16, ONA18}.

\begin{figure}[t]
\includegraphics[width=3.2in]{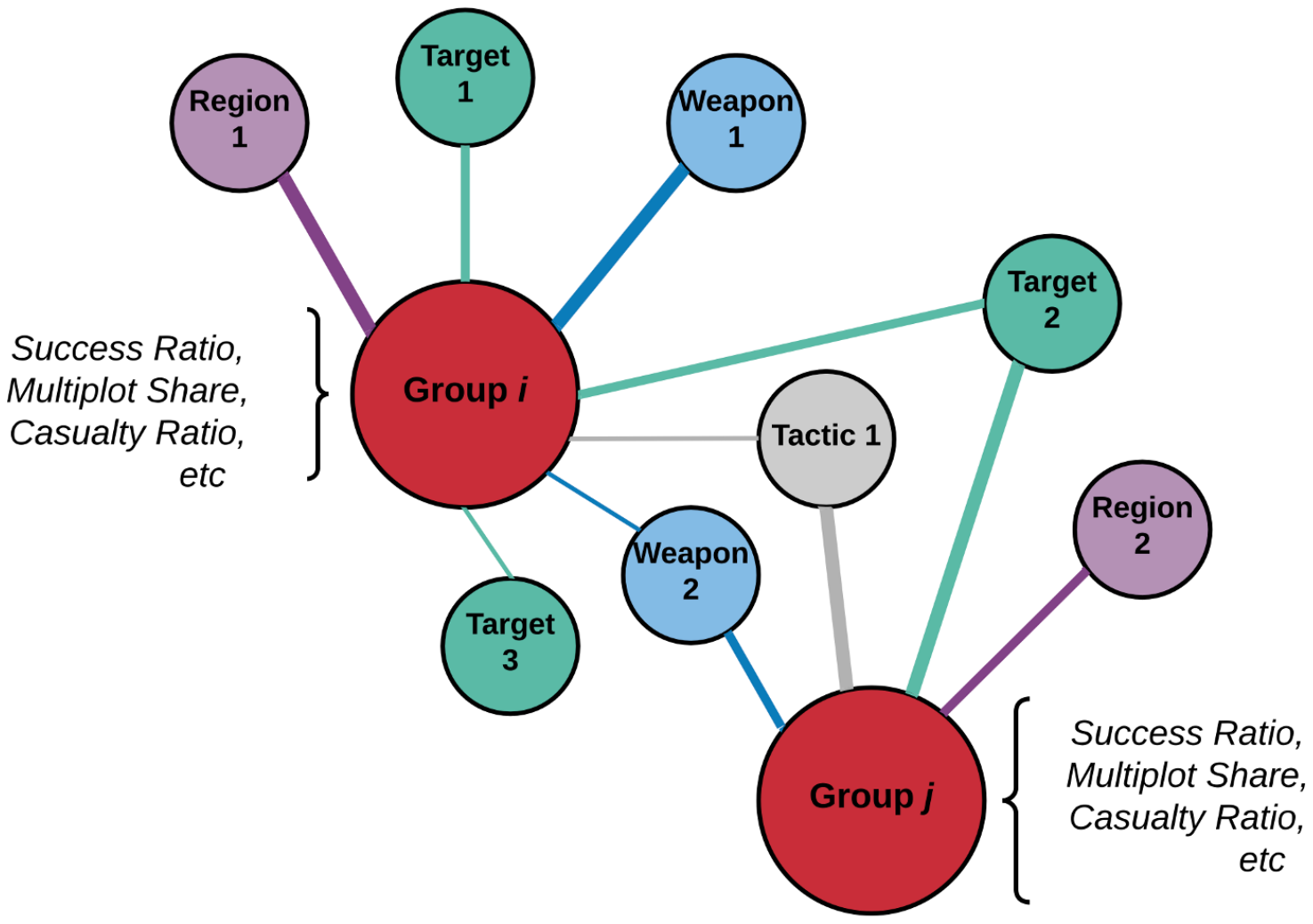}
\caption{
Schematic of a multi-partite graph constructed from a list of terrorist incidents. 
Red nodes $i,j$ represent perpetrator groups.  The numeric features intrinsic
to each group are assigned to the corresponding node and are denoted by 
$x_i^k$, where $i$ is the index labelling the terrorist group, and $k$ is the feature index.  
Here, we depict the success ratio ($k=1$), casualty ratio ($k=2$), and multiplot share ($k=3$)
as numeric features. 
An attack links the group that perpetrated it to all categorical features 
associated to the attack, including possible
regions (two purple nodes, $k=4,5$), targets (three green nodes, $k=6,7,8$), 
weapons (two blue nodes,  $k=9,10$), and tactics (one gray node,  $k=11$).  
The total number of features in this example is $K=11$. 
The connectivity of the graph generates the boolean variable $S_{i, j}^{(k)}$,
indicating whether groups $i$ and $j$ are linked to the same 
categorical feature $k$.
Since categorical feature nodes connect separately to terrorist group nodes, 
and not to each other, the multi-partite graph can be divided into four modes of bipartite graphs, 
one per categorical feature. 
Each has order $|V_k|$, for $k > 3$, given by the number of nodes in it.  For the region bipartite graph 
$|V_4| =|V_5| =  4$ (regions 1,2; groups $i,j$),  for the target bipartite graph
$|V_6| = |V_7| = |V_8| = 5$ (targets 1,2,3; groups $i,j$), for the weapons bipartite graph
$|V_9| = |V_{10}| = 4$ (weapons 1,2; groups $i,j$), 
and for the tactic bipartite graph $|V_{11}| =3$ (tactic 1; groups $i,j$).
Taken and adapted from Ref.\,\cite{CAM19b}.}
\label{FIG:multi-partite}
\end{figure}

One other issue is whether commonalities among terrorist groups can be found, and whether they 
influence one another in their {\textit {modus operandi}} or operational tactics. 
Campedelli et al. \cite{CAM19a, CAM19b} proposed a meta-graph method to classify 
terrorist groups by grouping their traits and strategic approaches. 
Using select GTD features, as illustrated in Fig.\,\ref{FIG:multi-partite},  a
multi-partite graph is constructed using categorical features (such as
operating regions, weapons of choice, target types, tactics, and ideologies) 
and numerical features (such as success rate, the percentage of suicide missions, 
fatality and casualty rates, number of targeted countries). 
Each of the $K$ features forms a bipartite graph with the set of terrorist groups, 
dividing the multi-partite graph into $N$ modes of bipartite graphs, 
The similarity between two terrorist groups $i$ and $j$ 
can be evaluated through Gower’s Similarity Coefficient \cite{GOW71}

\begin{equation}
   S_{i, j} = \frac{\sum_{k}  w_{k} \, S_{i,j}^{(k)}}{\sum_{k} w_{k}},  \qquad 1 \le k \le K  
     \label{eq:gower}. 
\end{equation}

\noindent
Here,  $S_{i,j}^{(k)}$ represents the similarity between groups $i$ and $j$ 
based on feature $k$, where $1 \le k \le K$,  
and $w_{k}$ is a weight coefficient. 
For categorical features $S_{i,j}^{(k)} = 1$ when 
both groups link to feature $k$, and $S_{i,j}^{(k)} = 0$ otherwise. 
For numerical features one defines 

\begin{equation}
   S_{i,j}^{(k)} = \frac{\left| x_{ik} - x_{jk} \right|}{r_k}, \label{eq:S_ijk}
\end{equation}

\noindent
where $x_{ik}$ is the numerical evaluation of feature $k$ for group $i$, 
and $r_k$ is the range of the values of feature $k$ across all groups. 
Finally, to evaluate the weight $w_k$ for each mode of the bipartite graph associated to feature $k$, an entropy-based weight 
coefficient is adopted \cite{CAM19b} so that 

\begin{equation}
   w_{k} = - \sum_{m} 
   {\frac{\lambda_m}{\left| V_k \right|} 
   \ln \frac{\lambda_m}{\left| V_k \right|}},  \qquad  1 \leq m  \leq {\left| V_k \right|}. 
   \label{eq:entropy-weight}
\end{equation}

\noindent
Here, $\lambda_m$ is the $m^{\textrm {th}}$ eigenvalue of the normalized Laplacian of 
the bipartite graph, and $\left| V_k \right|$ is its order.
Given that $\lambda_m \leq 2$ \cite{LIJ14} and
$|V_k| \geq 2$, as all bipartite graphs have at least two nodes, it follows that 
$\lambda_m  \leq |V_k|$, implying that
$w_k \geq 0$. Eq.\,\eqref{eq:entropy-weight}  assigns higher weights to graphs of more heterogeneous 
structures, highlighting modes that exhibit clusters.
The original multi-partite graph is then substituted by 
a latent similarity network $G$, where nodes represent terrorist groups, 
and edges are weighted by $S_{i,j}$ given in Eq.\,\eqref{eq:gower}. From $G$ one can
also determine the corresponding adjacency matrix $\mathbf{A}$. 
A k-nearest-neighbor (kNN) algorithm \cite{ZHA16} is then used to 
identify clusters in $G$, by iteratively maximizing 
the network modularity ${\mathrm {mod}}(G)$ given by

\begin{equation}
   \mathrm{mod}( G ) = \frac{1}{2m} \sum_{ij} 
   {\left[ A_{ij} - \frac{   \mathrm{deg}(i) \times   \mathrm{deg}(j)}{2m} \right]  \delta ( c_i, c_j ) }. 
   \label{eq:modularity}
\end{equation}

\noindent
Here,  $m$ is the total number of edges in $G$, 
$ \mathrm{deg}(i)$ is the degree of node $i$, 
and $c_i$ is the assigned cluster of node $i$ so that 
when $c_i = c_j$, $\delta (c_i, c_j) = 1$, 
otherwise, $\delta (c_i, c_j) = 0$. 
High modularity values imply that nodes in clusters of a graph 
are densely interconnected, while those that are part of different
clusters share fewer links \cite{CLA04, NEW10}.  Results show that Islamist terrorist groups 
tend to cluster separately from other ideologies, 
suggesting distinct behavioral patterns. 
Another interesting outcome is that far-left and far-right groups 
share similar cluster assignments, suggesting conformity in 
operational views despite their opposite ideologies. 
In addition, far-left groups  tend to form clusters 
with extremist environmental and animal rights groups, 
while far-right groups tend to side with ethno/nationalist 
and religious (non-Islamic) actors.  Similar
kNN methods were used to classify lexicons used to describe terrorist events 
so that motives, tactics, and targets of terrorist attacks could be grouped \cite{BRI22}.

\begin{figure}[t]
\includegraphics[width=4.7in]{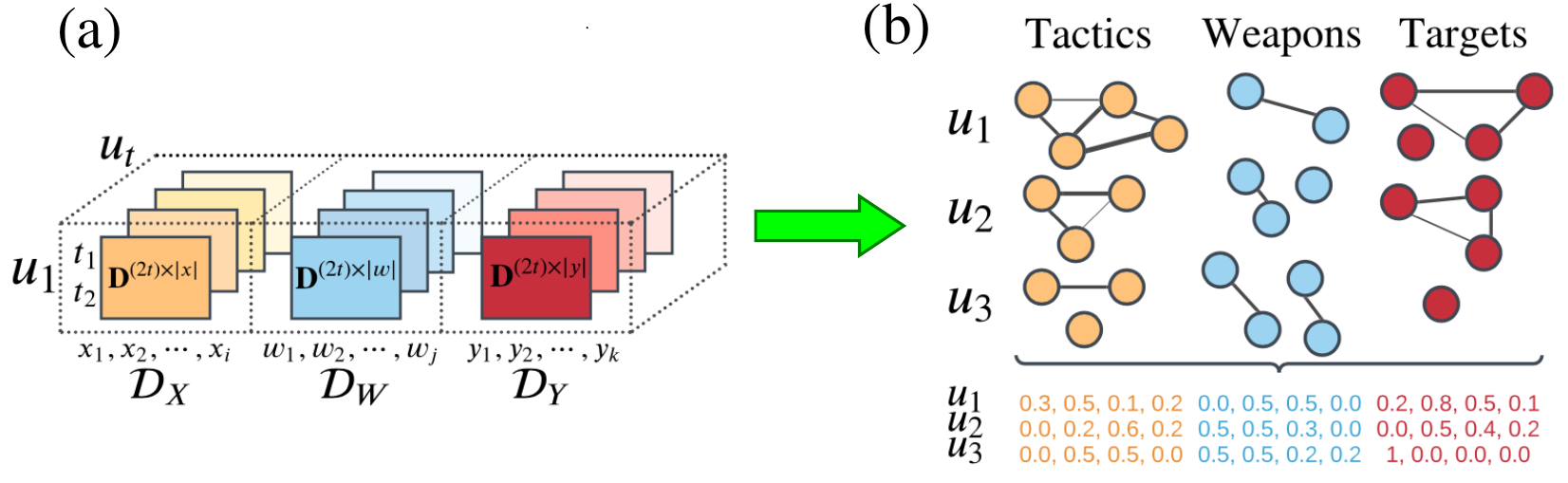}
\caption{
Illustration of data preprocessing converting a meta-graph data structure
on the left to a time series of normalized node centrality on the right.
Each time point $u$ is a two-day period $(t_1, t_2)$, with $u \in \left\{ u_1, ..., u_t \right\}$; 
there are $|x| =i$ tactics with ${\cal D}_X = \{ x_1, \dots x_i \}$, 
$|w| =j$ weapons with ${\cal D}_W = \{ w_1, \dots w_j \}$, and 
$|y| =k$ targets with ${\cal D}_Y = \{ y_1, \dots y_k \}$. 
(a)  The meta-graphs are abstract representations of a
full adjacency tensor of rank 4, constructed by mapping
all possible tactics, weapons, and targets used in terrorist events to nodes in a graph, and linking those
that appear together  in terrorist attacks at time $u$ as per the GTD. 
The procedure is described in the text.  The resulting meta-graphs are rank-2 matrices 
$\mathbf{D}^{2 \times |x|}$, $\mathbf{D}^{2 \times |w|}$, and $\mathbf{D}^{2 \times |y|}$
of dimension $2 \times |x|$, $2 \times |w|$, $2 \times |y|$, respectively whose
elements tally the occurrence of a specific tactic, weapon or target
in any attack at either of the $(t_1, t_2)$ times in $u$.  
(b) Using Eqs.\,\eqref{eq:centrality} and \eqref{eq:centrality_norm}, the meta-graphs
are reduced to a time series of normalized node centrality measures
$\psi_{\ell, \mathrm{Norm}} [ u ]$, representing the prevalence of each tactic, weapon,
or target at time $u$, and used as input for machine-learning algorithms.
In this example, each tactic, weapon,
or target feature has four nodes, i.e.,
$|x| = |w| = |y| = 4$, and the total dimension is
$|I| = |x|+ |w| + |y| = 12$.
The numeric sequence in the lower part of the panel represents
$\psi_{\ell, \mathrm{Norm}} [u]$ at time frames $u = u_1, u_2, u_3$,
for $1 \le \ell \le 12$. Since not all features may be represented
at a specific $u$, the number of nodes varies;
if node $\ell$ is absent at time $u$, then 
$\psi_{\ell, \mathrm{Norm}} (u) = 0$.
Taken and modified from Ref.\,\cite{CAM21b}.}
\label{FIG:meta-graph}
\end{figure}

The emergence of non-trivial clustering in operational characteristics suggests
that terrorist groups may imitate, or inspire, each other. 
Indeed, further studies with a time series of meta-graphs, 
called temporal meta-graphs, show that terrorist groups have 
become more similar over time \cite{CAM21a}. 
To the extent that later attacks are influenced by prior ones, 
machine learning techniques may be employed to forecast the occurrence
of future incidents.
Descriptors from three key features of terrorist attacks, namely  tactics
(${\cal D}_X$),
weap\-ons (${\cal D}_W$), and targets (${\cal D}_Y$), were taken
from the GTD to construct temporal meta-graphs \cite{CAM21b}.
The goal was to predict the risk of future attacks
based on operational characteristics of prior ones.
Here, it is assumed that there are $|x|= i$ possible tactics,
$|y| =j$ possible weapons and
$|w| = k$ possible targets, each represented by
a node in a graph. A series of
discrete time steps $u$ is also introduced, each consisting of two consecutive days
to avoid an overly sparse representation.
A weighted hyper-graph is generated
at each $u$ by linking tactic, weapon, and target nodes
that manifest together in one or more attacks. The weight of the link is given by the number of attacks
(within $u$) where the same operational characteristics are used; if there are none, the
link is assigned value zero.
The weighted hyper-graphs give rise to an adjacency tensor of rank 4,
with elements labeled by $u, x, w, y$, where $u \in \{u_1, \dots u_t \}$ is the discrete time step and
$x \in \left\{x_1, \dots x_i\right\}$, $w \in \left\{w_1, ..., w_j\right\}$, and
$y \in \left\{y_1, ..., y_k\right\}$ represent all possible operational tactics, weapons, and targets, respectively.
To further simplify the data structure,
the adjacency tensor is collapsed into three rank-2 matrices
$\mathbf{D}^{(2t) \times |z|}$
 where $|z| \in \{|x|,|w|,|y| \}$, one
for each of the three feature dimensions, as illustrated in Fig.\,\ref{FIG:meta-graph}\,(a).
The dimension of each matrix is $(2t) \times |z|$ since each time step $u$ is made of two days, and there are $t$ time steps.
Each $\mathbf{D}^{(2t) \times |z|}$ matrix is then divided into $t$ slices,
defining $t$ matrices of dimension $2 \times |z|$, denoted by $\mathbf{D}^{(2) \times |z|} [u]$
and referred to as ``meta-graphs."  Its matrix elements are the degrees of the corresponding node in the hyper-graph
on a given day,  and represent how many times the specific tactic $x$, weapon $w$, or target $y$
was associated with any other on that day through a common terrorist event. 
The meta-graphs are then  concatenated
horizontally into a single matrix $\mathbf{D} [u]$, of dimension $2 \times |I|$,
where $ | I | =  |x|+ |y| + |w|$.
Finally, $\mathbf{D} [u]$ is used to determine the frequency of each
operational characteristic $\ell \in \{1, \dots, |I|\}$ among terrorist events at time $u$.  This is done by
introducing a centrality metric $\psi_{\ell} [ u ]$ defined as

\begin{equation}
   \psi_{\ell} [ u ] = \sum_{m = 1, m \ne {\ell}}^{\left| I \right|} \mathbf{G}_{\ell, m}[ u ], \label{eq:centrality}
\end{equation}

\noindent
where $\mathbf{G} [ u ] \equiv \mathbf{D}[ u ]^{\mathrm{T}} \mathbf{D} [ u ]$
is a $\left| I \right| \times \left| I \right|$ matrix at time $u$.
Before being applied to machine-learning models, $\psi_{\ell}$ is further normalized

\begin{equation}
   \hat{\psi}_{\ell} [ u ] = \frac{\psi_{\ell}[ u ]}{\max_{\ell  \in I} \psi_{\ell} [ u ]}. \label{eq:centrality_norm}
\end{equation}

\noindent
Fig.\,\ref{FIG:meta-graph}\,(b) shows an example of such preprocessed data.
Of the terrorist events occurring in Iraq and Afghanistan between 2001 and 2018, 
the first $70 \%$ were used for training (about 12.5 years), 
the next $20 \%$ for validation (about 3.5 years), 
and the last $10 \%$ for testing (about two years). 
A variety of machine-learning algorithms and their combinations were
tested, including a traditional feedforward neural network (FNN), 
which treats each data input as an independent event, 
a long short‐term memory network (LSTM), 
which is a recurrent neural network useful for time-series data, 
and a convolution neural network (CNN) typically employed for grid data. 
The performance of these algorithms was evaluated on the basis of 
three distinct criteria: the standard mean-squared error (MSE) to minimize discrepancies
between data and predictions, an element-wise accuracy (EWA) for capturing the top two targets, 
and a set-wise accuracy (SWA) for predicting the list of targets. 
All machine-learning models 
(FNN, LSTM, CNN) that used data 
preprocessed through meta-graphs were found to yield superior predictions 
compared with baseline models that only use the number of attacks. 
This suggests that connections between features
(such as the ones constructed through the
meta-graphs shown in Fig.\,\ref{FIG:meta-graph}), and not just their frequency, 
hold valid information useful in forecasting future terrorist attacks. 
Furthermore, the LSTM network outperforms other protocols, 
indicating that past attacks affect the unfolding of future ones. 

Olabanjo et al. \cite{OLA21} developed selection models to filter among all the features listed 
by the GTD, and found that a hybridized model combining the Chi-square and the Information
Gain models can best predict the occurrence of terrorist attacks.
Other machine-learning attempts at forecasting terrorist attacks (time, location, lethality), 
mostly using the GTD alone or in conjunction with other datasets, 
include using hidden Markov models \cite{PET13, SHE07}, regression analyses \cite{PYT19, YAN19}, 
classifiers such as kNN, RF, support vector machine (SVM) \cite{GOH14, TOL15, DIN17, KRI22, PYT21},
deep neural networks such as recurrent neural networks (RNN) and FNN \cite{DIN17, UDD20}, 
and other methods \cite{BAS18, HUA20, BUF22, PAN23}.   

Aside from predicting potential targets or locations of future terrorist events, 
ma\-chine-learning methods can also be used to learn about 
risk factors associated to attacks.
By applying the Knox test to over 200,000 GTD entries, 
Hunt et al. \cite{HUN24} identified the 30 major terrorist groups 
responsible for 99$\%$ of terrorist events worldwide between 1970 and 2020. 
This analysis showed that  28 out of the 30 terrorist groups exhibit 
significant near-repeat tendencies, expressed by a Knox ratio greater than 
1.2 over several spatio-temporal thresholds.  All events are then marked depending on whether they
are part of an identified cluster or not.  Several machine-learning models are then used to 
predict the ones that are marked as being part of a cluster drawing on 43 select features from the 
GTD.  Using the conventional 80-20 training-test split ratio, it is observed
the Random Forest (RF) model attains the best performance, in terms of precision and sensitivity.
The method of permutation importance is finally implemented to the RF 
to assess the relative importance of each feature contributing to near-repeat attacks. 
Here, the values of a selected feature are randomly shuffled within the data 
and ensuing predictions are compared with the original ones.
If the selected feature is relatively more important, 
random permutations would result in poorer predictions. 
\begin{figure}[t]

\includegraphics[width=4.7in]{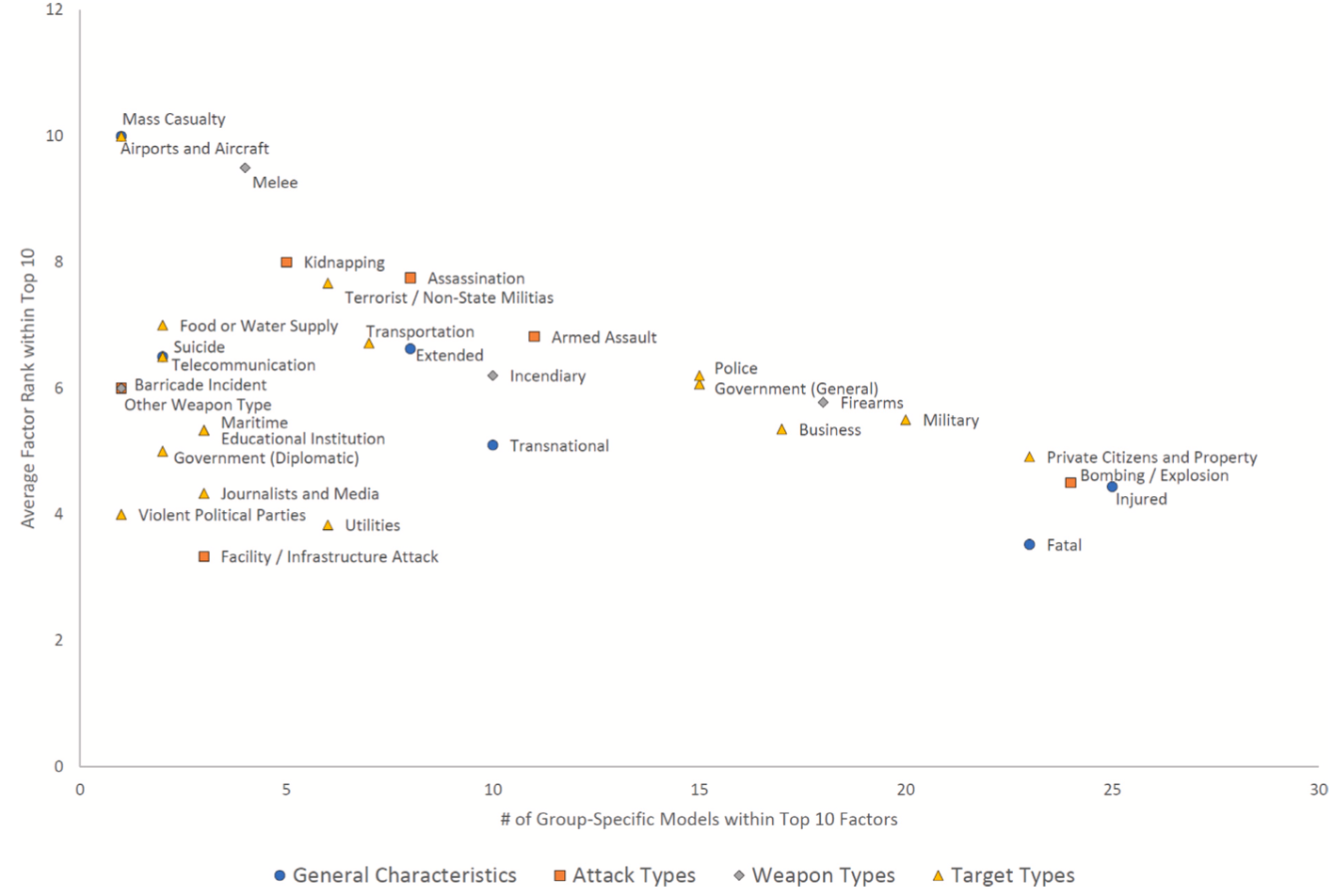}
\caption{Risk factors of near-repeat attacks estimated from the activity of major terrorist groups as per the GTD. 
The horizontal axis represents the number of terrorist groups (out of the 28 that exhibit near-repeat clustering)
or which the specified feature in the graph is among the top ten risk factors. 
The vertical axis represents the average rank of the given feature among the top ten in descending order, 
with 1 the highest.  Thus, the most common and highest ranking risk
targets, weapons, attack types are private citizens and property, firearms,  bombing and explosions,
respectively.  The risk is also highest for fatalities and injuries. 
Taken from Ref.\,\cite{HUN24}.}
\label{FIG:Hunt24}
\end{figure}
Findings are summarized in Fig.\,\ref{FIG:Hunt24}.  The most common and highest ranking 
risk factors are for fatalities and injuries,  bombing and explosion tactics, targets of private citizens and property, 
and firearm weapons.

\section{Summary and conclusions}
\label{conclusions}

In this brief review, we surveyed a few mathematical models that studied
radicalization trajectories, how extremist groups organize, and 
patterns of terrorist attacks. These quantitative models
are built from findings derived from the sociological literature, case studies, interviews and surveys, 
counter-terror reports, political science theories, data analysis
and empirical observations, which are all limited by the covert nature of terrorist organizations
and by the difficulty in conducting controlled tests. 
Much work is still necessary for a more complete understanding of the phenomena at hand, from all angles. 
For example, while data and interviews may offer an 
\textit{a posteriori} understanding of why or how some individuals 
radicalize or become terrorists, they cannot
explain why others exposed to the same environment do not. Most researchers show that early intervention is the 
optimal way to avoid the emergence of large scale radical factions and operative terrorists,  
but the legal implications of pre-emptive, invasive policy making must be factored in. 
Other advancements include a more rigorous comparative analysis 
to characterize terrorism in different historical and sociopolitical 
contexts, 
better distinction between 
online and offline radicalization, the emergence of lone wolves as opposed to group dynamics.
Developing novel methods, including machine learning approaches, to dissect information from material that terrorists themselves
divulge, in the form of videos and other internet content would offer better insight into their \textit{modus
pensandi}. Most work has focused on traditional attacks, but new forms of terrorism that would elicit different
societal and governmental responses should also be investigated, 
such as the use of chemical, biological, radiological weapons.
How to prevent the weaponization of trucks, airplanes, drinks or other materials is also of great concern.  
Exploring the role of the media, how to best use the internet as a counter-terrorism tool, 
how to create new channels of communications between counter-terror agencies
across geographical boundaries, would help shape innovative intervention methods. 
Also, one must be aware that tactics, players, allegiances, and
technology can change rapidly so that current findings and models
may not be robust to the passing of time, or universally applicable.
Finally, developing prevention programs, finding potential pathways out of radicalization,
are also important issues. 

Any attempt to answer these questions must involve experts and practitioners
from many disciplines, that include the humanities, social sciences, ethic studies, 
economics, computer and quantitative sciences.
Mathematical modeling has great potential, 
as it can help build predictive models where the consequences of different inputs, strategies, 
and actors may be examined, where costs and information constraints may be quantified.
However, all players, including on-the-ground communities, strategies and mechanisms, must be cogently included.
It is thus important that a common language between social and 
mathematical scientists be developed, that academic and practical 
opportunities for dialogue be created, and that each community actively 
engage with the other to avoid oversimplifications.
Similarly, data that used to train machine learning methods or verify predictions
must be accurately collected to avoid biases or profiling, while ensuring privacy 
especially when it comes to individual behaviors.

Great care and clarity must be exercised in translating any new finding into actual policy-making, including
ethical considerations, robust validation, and monitoring of results.

\begin{ack}
We thank Profs. Tom Chou, Lucas B\"ottcher, Serge Galam for insightful comments. 
\end{ack}

\begin{funding}
This work was supported by the Army Research Office through W911NF-23-1-0129 
and by the National Science Foundation through grant MRI-2320846.

\end{funding}










\end{document}